\documentclass[%
reprint,
amsmath,amssymb,
aps
]{revtex4-1}
\usepackage{url}
\usepackage[colorlinks=true, linkcolor=blue,urlcolor=blue,anchorcolor=blue,citecolor=blue,bookmarksnumbered]{hyperref}
\usepackage{graphicx}
\usepackage{dcolumn}
\usepackage{bm}

\begin{document}

\title{Systematic error tolerant multiqubit holonomic entangling gates}
\author{Jin-Lei Wu$^{1}$}\author{Yan Wang$^{1}$}\author{Jin-Xuan Han$^{1}$}\author{Yongyuan Jiang$^{1}$}\author{Jie Song$^{1}$}\email[]{jsong@hit.edu.cn}\author{Yan Xia$^{4}$}\author{Shi-Lei Su$^{2}$}\email[]{slsu@zzu.edu.cn}\author{Weibin Li$^{3}$}\email[]{weibin.li@nottingham.ac.uk}
\affiliation{$^{1}$School of Physics, Harbin Institute of Technology, Harbin 150001, China\\
	$^{2}$School of Physics and Microelectronics, Zhengzhou University, Zhengzhou 450001, China\\
	$^{3}$School of Physics and Astronomy, University of Nottingham, Nottingham NG7 2RD, United Kingdom\\
	$^{4}$Department of Physics, Fuzhou University, Fuzhou 350002, China}

\begin{abstract}
Quantum holonomic gates hold built-in resilience to local noises and provide a promising approach for implementing fault-tolerant quantum computation. We propose to realize high-fidelity holonomic $(N+1)$-qubit controlled gates using Rydberg atoms confined in optical arrays or superconducting circuits. We identify the scheme, deduce the effective multi-body Hamiltonian, and determine the working condition of the multiqubit gate. Uniquely, the multiqubit gate is immune to systematic errors, i.e., laser parameter fluctuations and motional dephasing, as the $N$ control atoms largely remain in the much stable qubit space during the operation. We show that $C_N$-NOT gates can reach same level of fidelity at a given gate time for $N\leq5$ under a suitable choice of parameters, and the gate tolerance against errors in systematic parameters can be further enhanced through optimal pulse engineering. In case of Rydberg atoms, the proposed protocol is intrinsically different from typical schemes based on Rydberg blockade or antiblockade.  Our study paves a new route to build robust multiqubit gates with Rydberg atoms trapped in optical arrays or with superconducting circuits. It contributes to current efforts in developing scalable quantum computation with trapped atoms and fabricable superconducting devices.
\end{abstract}
\maketitle

\section{Introduction}
Faithfully and efficiently implementing quantum gates among multiple qubits is of central task in building near-term quantum computing systems~\cite{Nielsen}. Yet typical gate errors are caused by decoherence due to qubit-environment coupling, and systematic errors due to imperfect state preparation and operation~\cite{Galindo2002,Nigg2014}. Inspired by the intrinsic resilience to environmental noises of geometric phases~\cite{Berry1984,Leek2007,Filipp2009}, adiabatic paradigms~\cite{Zanardi1999,Duan2001,Wu2005} of holonomic quantum computation~(HQC) based on non-Abelian geometric phases~\cite{Wilczek1984} have been designated, following which shortcut to adiabatic~\cite{Chen2010,Chen2014,STA2019,Song2016,JZhang2015,Chen2020} and nonadiabatic HQC~(NHQC)~\cite{Sjoqvist2012,Xu2012,Hong2018,Liu2019,Zhao2020,Ota2009,Kondo2011,Merrill2014} have been proposed and demonstrated experimentally with single- and two-qubit gates~\cite{Abdumalikov2013,Feng2013,Zu2014,Arroyo2014,Long2017,Sekiguchi2017,Zhou2017,Xue2018,Zhang2019,Yan2019}. Efforts have been spent to combine HQC with decoherence-free subspace to protect qubits from noises~\cite{Wu2005,Xu2012} or with error-correcting codes to eliminate qubit errors actively~\cite{Oreshkov2009,JZhang2018,CWu2020,Chen2021}. With these developments, the gate errors are mainly affected by the imperfect control parameter~\cite{Liu2019,Zheng2016,Liu2021}. It can limit the overall fidelity of lengthy algorithms in quantum computation~\cite{Maslov03,Mueller09,Isenhower11,Martinez16,Khazali2020,Rasmussen20,Yin2020,Young2021} and simulation~\cite{Weimer10,Keesling19,Gambatta20a,Gambatta20b}, as typically the HQC is implemented with concatenating multiple, universal single- and two-qubit gates.

\begin{figure}[b]\centering
	\includegraphics[width=0.68\linewidth]{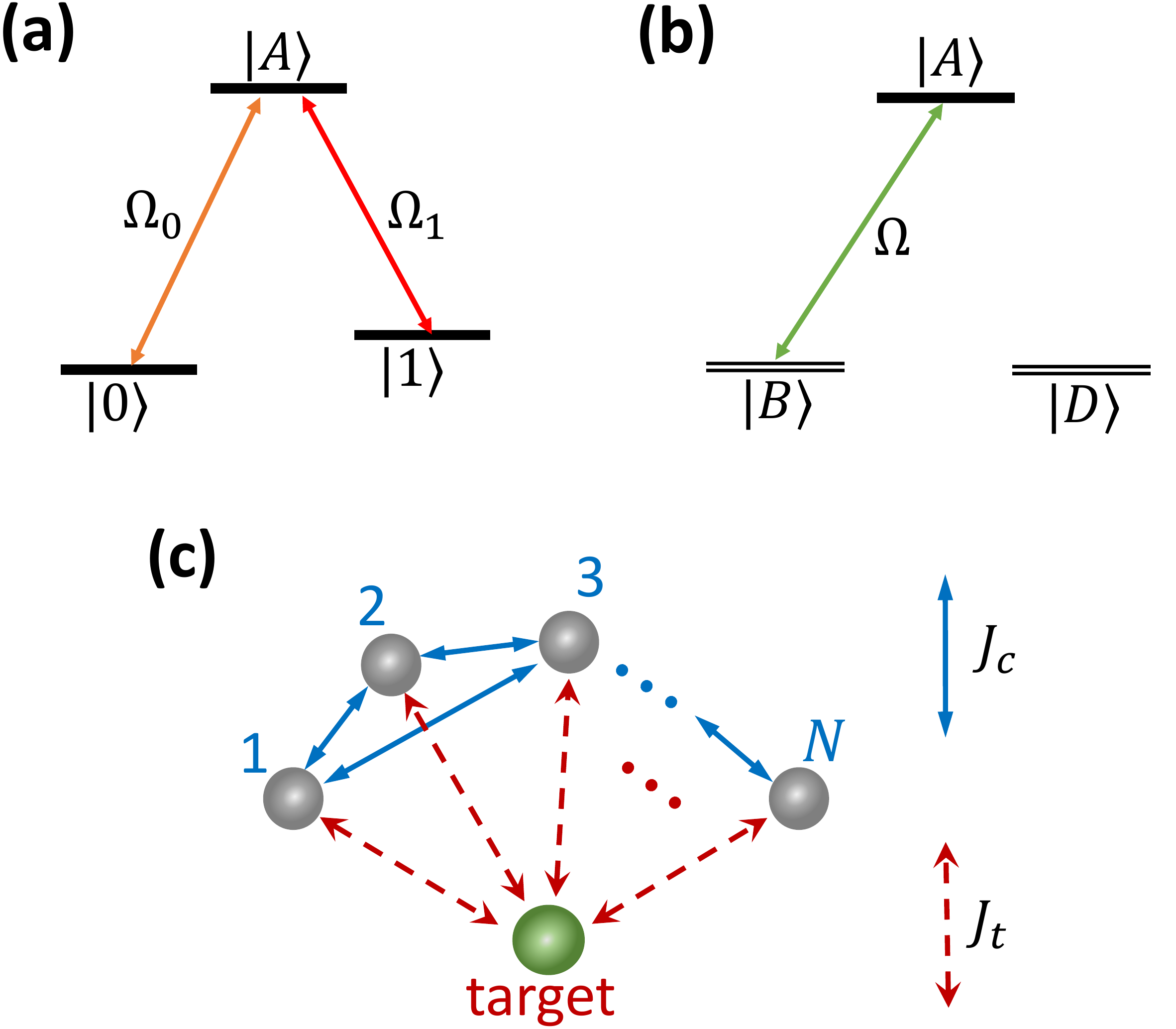}
	\caption{(a)~Level diagram. Two low-lying states $|0\rangle$ and $|1\rangle$ couple resonantly to an auxiliary state $|A\rangle$ with Rabi frequencies $\Omega_0$ and $\Omega_1$. (b)~State $|A\rangle$ resonantly couples to dressed state $|B\rangle$ (Rabi frequency $\sqrt{\Omega_{0}^2+\Omega_{1}^2}$) but not to dressed state $|D\rangle$. (c)~Two-body Ising (density-density) interaction between the control qubits (strength $J_c$), and between the control and target qubits (strength $J_t$).}\label{f1}
\end{figure}

In this work, we propose a one-step approach for implementing $(N+1)$-qubit holonomic gates~[$(N+1)$-QHG] in a many-body model where qubits interacting with each other by the Ising interaction.
Rydberg atoms, as natural qubits with perfect identity, are a promising platform for realizing the Ising model by means of van der Waals or dipole-dipole interactions among atoms, and recent experiments have shown cryogenic atom cooling, large-probability atom trapping and loading, defect-free arrangement of atom arrays, and high-fidelity single-atom manipulations~\cite{Ebadi2021,Scholl2021,Bluvstein2021}. With Rydberg atoms in an optical array~\cite{Nogrette2014,Wang2015,Endres2016,Barredo2016,Wang2016,Kumar2018,Barredo2018}, we identify an unconventional regime where the Rydberg antiblockade condition is employed but the blockade phenomenon emerges. In this regime, all $N$ control atoms remain unexcited in the gate. We obtain the parameter region where increasing $N$ does not demote the driving strength of the effective system. Importantly, the error sensitivity to fluctuations in laser parameters and interatomic distances can be largely suppressed through optimal pulse engineering~\cite{Ruschhaupt2012,Daems2013,JZhang2017}. Moreover, we show that the multiqubit holonomic gates can be realized alternatively with superconducting circuits. Our study paves a route to achieve optimized holonomic quantum computation with strongly interacting Rydberg atoms and superconducting circuits, and might find applications in quantum computation and simulations based on robust multiqubit gates.

This paper is organized as follows: in section~\ref{sec2}, we introduce the many-body model and show the deviation of effective Hamiltonian for implementing $(N+1)$-QHG. In section~\ref{sec3}, we describe the realization of the many-body model with a Rydberg atom array and show simulations for implementing C$_N$-NOT gates of NHQC. In section~\ref{sec4}, we give an optimal pulse scheme for robust $(N+1)$-QHG and identify the excellent robustness of the scheme. Finally, a conclusion is given in section~\ref{sec5}.

\begin{figure}[b]\centering
	\includegraphics[width=\linewidth]{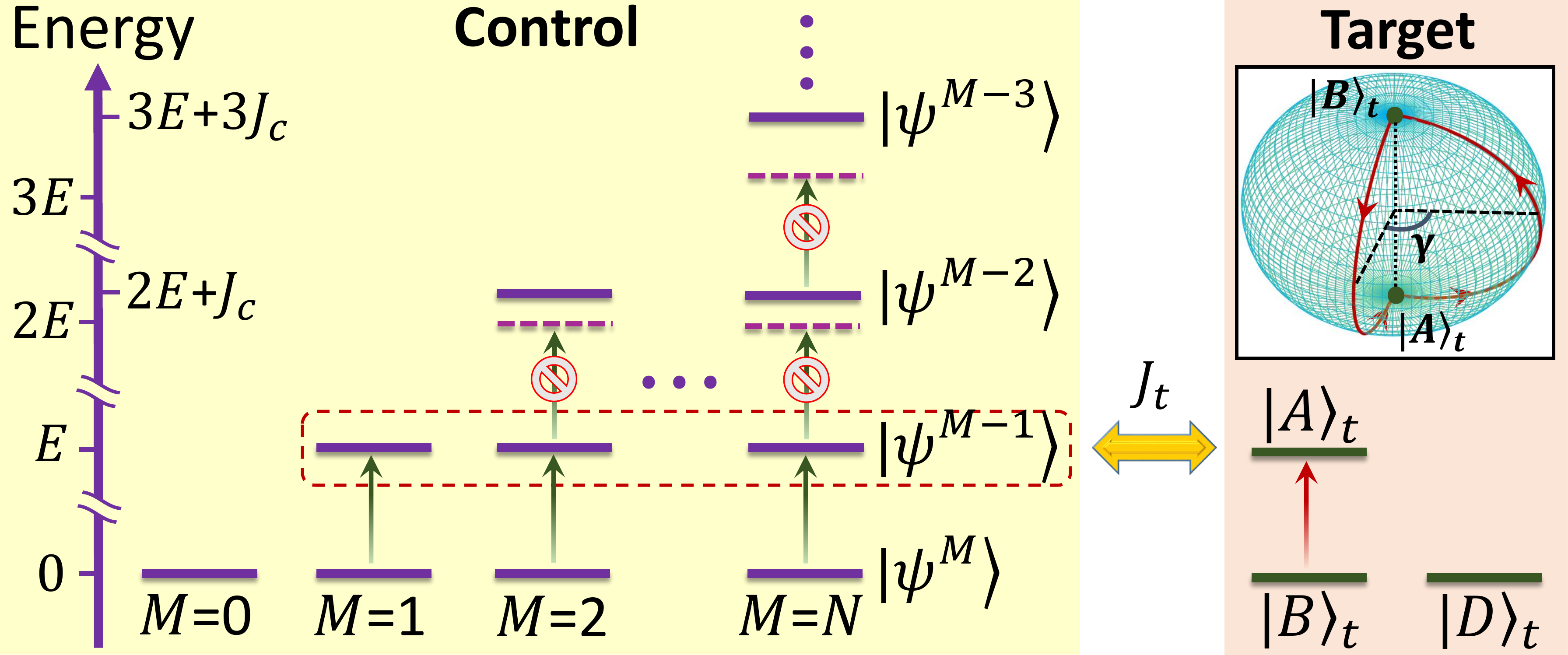}
	\caption{Collective states of the control qubits (left panel), $M$ denoting the number of qubits initially in state $|B\rangle$. Due to the strong Ising interaction, the collective states with multiple qubits in state $|A\rangle$ are prohibited from excitation. When $J_t$ is large, a geometric gate can be realized through one step operation. The gate trajectory on the Bloch sphere is shown on the right panel.}\label{f2}
\end{figure}
\section{Many-body model}\label{sec2}
We first present an $(N+1)$-qubit model that is independent of physical systems. Each qubit consists of two computational states $|\alpha'\rangle~(\alpha'=0,1)$ and an auxiliary state $|A\rangle$, as shown in Fig.~\ref{f1}(a). States $|\alpha'\rangle$ and $|A\rangle$ are coupled resonantly with Rabi frequencies $\Omega_{\alpha'}e^{i\varphi_{\alpha'}}$ ($\Omega_{\alpha'}$ and $\varphi_{\alpha'}$ being the amplitude and phase). By introducing dressed states $|B\rangle\equiv\sin(\theta/2)|0\rangle-\cos(\theta/2)e^{i\phi} |1\rangle$ and $|D\rangle\equiv\cos(\theta/2)|0\rangle+\sin(\theta/2)e^{i\phi} |1\rangle$ with $\theta\equiv2\arctan(\Omega_{0}/\Omega_{1})$ being constant and $\phi\equiv\varphi_{1}-\varphi_{0}-\pi$, only states $|B\rangle$ and  $|A\rangle$ are coupled with Rabi frequency $\Omega=\sqrt{\Omega_0^2+\Omega_1^2}$ [see Fig.~\ref{f1}(b)], described by Hamiltonians~($\hbar\equiv1$) $\hat{H}_{c}= \frac{\Omega_{c}}2\sum_{j=1}^N \hat X_j$ for the control qubits and $\hat{H}_{t}= \frac{\Omega_{t}}2\hat X_{t}$ for the target qubit, with $\hat X=|B\rangle\langle A|+|A\rangle\langle B|$~(hereinafter we label the control and target qubits with subscripts $c$ and $t$, respectively).
To be convenient, all control-control~(control-target) interactions are assumed to have equal strength $J_c$~($J_t$), as shown in Fig.~\ref{f1}(c). Such assumption, however, is not crucial, as we will illustrate with Rydberg atoms later. This assumption leads to a compact form of the interaction Hamiltonian $\hat{H}_{\rm Ising}= \sum_{j>k=1}^NJ_c{\hat n}_j{\hat n}_k+J_t{\hat n}_j{\hat n}_{t}$, where ${\hat n}_j=|A\rangle_j\langle A|$ is the number operator. The total Hamiltonian is given by $\hat{H} = \hat{H}_c + \hat{H}_t + \hat{H}_{\rm Ising}$.

\subsection{Hamiltonian with Dicke states}
To reveal the working mechanism of the multiqubit gate, we introduce collective states of the control qubits~(similar to Dicke states of two-level atoms~\cite{Dicke1954}) $|\psi^K\rangle\equiv|D\rangle^{N-M}|B\rangle^K|A\rangle^{M-K}$ consisting of $(N-M)$, $K$, and ${(M-K)}$ qubits in states $|D\rangle$, $|B\rangle$, and $|A\rangle$, respectively ($0\leq K\leq M\leq N$)~\cite{Moller2008,Rasmussen20}.
The states $|\psi^K\rangle$ and $|\psi^{K-1}\rangle$~($K\ge 1$) are coupled with collective Rabi frequency $\sqrt{K'}\Omega_c$ with $K'={K(M-K+1)}$~\cite{Noguchi2012,YFXiao2007,Jones2009,Lettner2011,Bernien2017,Browaeys2020,Qin2021}~(see~Appendix~\ref{Appendix_A} for details). Then, the composite basis of the collective control and target qubits is denoted with $|\chi_1^K\rangle\equiv|\psi^K\rangle|B\rangle_{t}$, $|\chi_2^K\rangle\equiv|\psi^K\rangle|A\rangle_{t}$, $|\chi_3^K\rangle\equiv|\psi^{K-1}\rangle|B\rangle_{t}$, and $|\chi_4^K\rangle\equiv|\psi^{K-1}\rangle|A\rangle_{t}$, in which there are $N_A=M-K$ ($N_A=M-K+1$) control qubits in state $|A\rangle$ for $|\chi_{1,2}^K\rangle$ ($|\chi_{3,4}^K\rangle$). The Ising energy in state $|\chi_p^K\rangle$ is $E_p^{(K)}=C_{N_A}^2J_c +\xi N_AJ_t$  with $\xi =0$ ($\xi=1$) for $p=1,3$ ($p=2,4$), $C_{N_A}^2$ being the binomial coefficient. Consequently, we can rewrite the Hamiltonian $\hat{H}$ with the composite states~(the derivation is given in Appendix~\ref{Appendix_A})
\begin{eqnarray}\label{e1}
	\hat{H}&=& \sum_{M=0}^N\sum_{K=0}^M\Big[\frac{\sqrt{K'} \Omega_{c}}{2}\Big({\hat S}_{31}^{(K)} e^{it\Delta_{31}^{(K)}}+{\hat S}_{42}^{(K)} e^{it\Delta_{42}^{(K)}}\Big)\nonumber\\
	&&+\frac{\Omega_{t}}{2}\Big({\hat S}_{21}^{(K)}e^{it\Delta_{21}^{{(K)}}}+{\hat S}_{43}^{(K)} e^{it\Delta_{43}^{(K)}}\Big)\Big]+{\rm H.c.},
\end{eqnarray}
where ${\hat S}_{pq}^{(K)}=|\chi_p^K\rangle\langle\chi_q^K|$ and $\Delta_{pq}^{(K)}=E_{p}^{(K)}-E_{q}^{(K)}$ are the collective transition operator and detuning, respectively. 

\subsection{Effective Hamiltonian}
	We introduce an amplitude-modulated field $\Omega_{c} =\bar{\Omega}_c\cos\omega t$, where $\bar{\Omega}_c$ is the amplitude and $\omega$ is the modulation frequency. $\omega$ is set with the same order of magnitude as the smallest of nonzero ones in $\{\Delta_{pq}^{(K)}\}$. In the strong interaction regime we set $|\omega|,|\Delta_{pq}^{(K)}|\gg|\bar{\Omega}_{c}|,|{\Omega}_t|$ when $K\neq M$, only the terms of $K=M,(M-1)$ remain in consideration, with $E_{1}^{(M)}=E_{2}^{(M)}=E_{3}^{(M)}=E_{1}^{(M-1)}=0$, $E_{4}^{(M)}=E_{2}^{(M-1)}=J_t$, $E_{3}^{(M-1)}=J_c$, and $E_{4}^{(M-1)}=J_c+2J_{t}$. The Hamiltonian~(\ref{e1}) becomes ${\hat H}\approx {\hat H}_{13}+{\hat H}_{124}$,
\begin{eqnarray}\label{S3}
	{\hat H}_{13}
	&=&\sum_{M=1}^N\frac{\sqrt M \bar{\Omega}_c}{4}|\chi_3^M\rangle\langle\chi_1^M|(e^{i\omega t}+e^{-i\omega t})\nonumber\\
	&&+\sum_{M=2}^N\frac{\sqrt {M-1} \bar{\Omega}_c}{4}|\chi_3^{M-1}\rangle\langle\chi_3^M|e^{it(J_c-\omega)}+{\rm H.c.},\nonumber\\
	\cr {\hat H}_{124}&=&\sum_{M=0}^N\Big[\frac{\sqrt M \bar{\Omega}_c}{4}{\hat S}_{42}^{(M)} e^{it(J_{t}-\omega)}
	+\frac{\Omega_{t}}{2}{\hat S}_{21}^{(M)}\Big]+{\rm H.c.},\nonumber\\
\end{eqnarray}
where terms with highly frequent oscillations have been neglected.

From the expression of ${\hat H}_{13}$ we know that even though the second term can be resonant, ${\hat H}_{13}$ works insignificantly because the first term of ${\hat H}_{13}$ can hardly induce a transition from $|\chi_1^M\rangle$ to $|\chi_3^M\rangle$ when $\omega\gg\sqrt N\bar{\Omega}_c/4$. Therefore, we obtain an effective Hamiltonian 
\begin{equation}\label{S4}
	{\hat H}'=\sum_{M=0}^N\Big[\frac{\sqrt M \bar{\Omega}_c}{4}{\hat S}_{42}^{(M)} e^{it(J_{t}-\omega)}
	+\frac{\Omega_{t}}{2}{\hat S}_{21}^{(M)}\Big]+{\rm H.c.},
\end{equation}
which shows that the detuning $\omega$ can exactly compensate for the control-target interaction $J_{t}$.
On the other hand, from $\hat H_{13}$ we learn that due to the large detuning a strong control-control interaction $J_c$ is unnecessary for suppressing doubly- or multiply-populated auxiliary states. Therefore, the Ising interaction takes place only between the target qubit and collective state with a single auxiliary state. Such a selective many-body interaction can be visualized in Fig.~\ref{f2} and is the key to realizing controlled multiqubit gate operation.

\begin{figure}[t]\centering
	\includegraphics[width=0.8\linewidth]{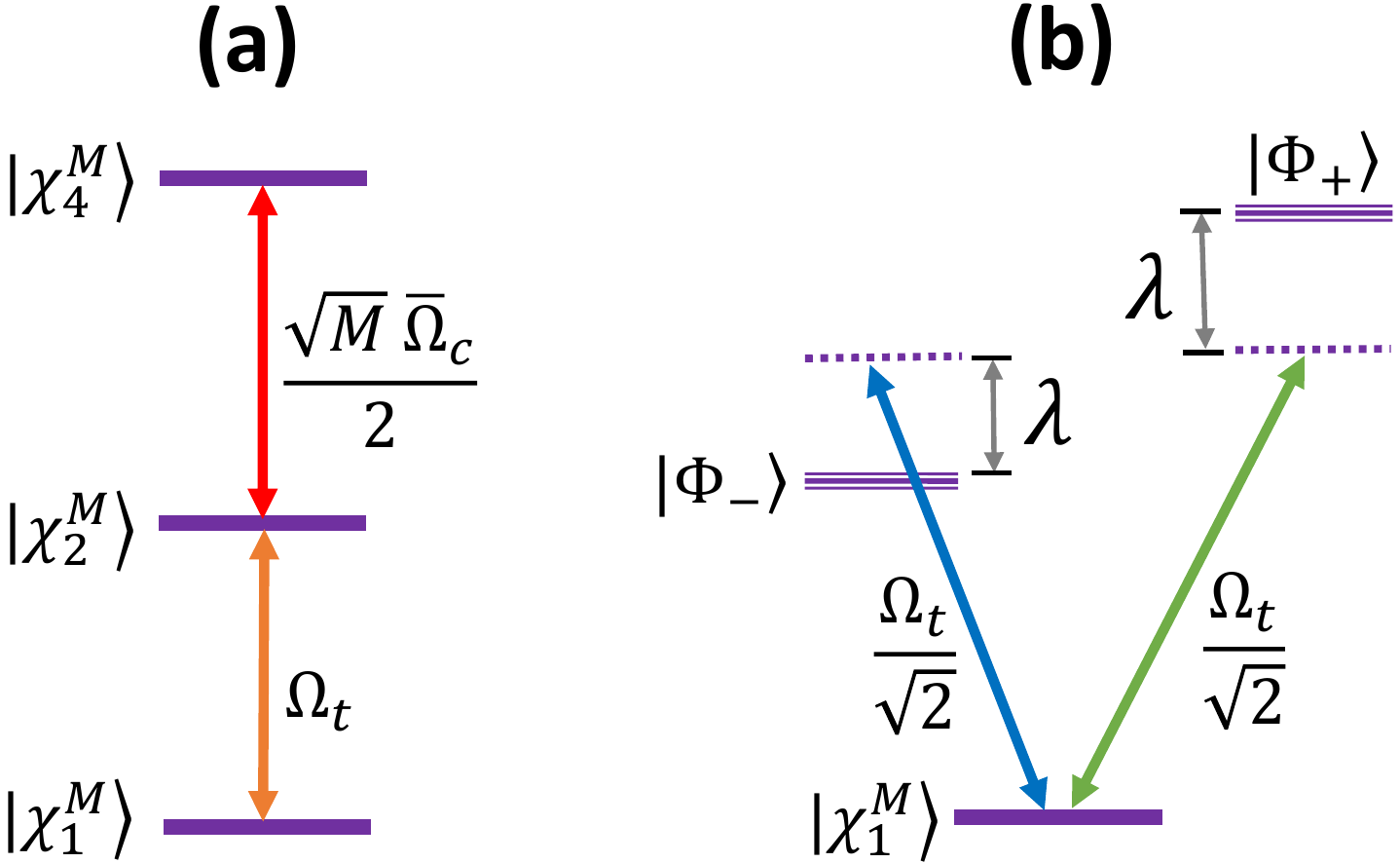}
	\caption{{(a)}~Three-level interaction with $M\neq0$ in Eq.~(\ref{S5}). {(b)}~Three-level interaction represented on the dressed-state basis.}\label{f3}
\end{figure}

Further we choose $\omega=J_t$, so the interference between the amplitude modulation of the drive on the control qubits and the control-target interaction gives rise to an effective Hamiltonian
\begin{equation}\label{S5}
	\hat{\mathcal{H}}_{\rm eff}=\sum_{M=0}^N \Big[\frac{\sqrt M \bar{\Omega}_c}{4}\Big({\hat S}_{24}^{(M)}+{\hat S}_{42}^{(M)}\Big)+\frac{\Omega_{t}}{2}\Big({\hat S}_{12}^{(M)}+{\hat S}_{21}^{(M)}\Big)\Big].
\end{equation}
This Hamiltonian involves a series of $M$-dependent~($M\neq 0$) three-level systems containing transitions $|\chi_{1}^M\rangle\leftrightarrow|\chi_{2}^M\rangle$ and $|\chi_{2}^M\rangle\leftrightarrow|\chi_{4}^M\rangle$, respectively, with Rabi frequencies $\Omega_{t}$ and ${\sqrt M \bar{\Omega}_c}/{2}$ as shown in Fig.~\ref{f3}{(a)}. We consider the diagonalization of the interaction $\hat{h}_1={\sqrt M \bar{\Omega}_c}[{\hat S}_{24}^{(M)}+{\hat S}_{42}^{(M)}]/{4}\rightarrow\hat{h}'_1= {\sqrt M \bar{\Omega}_c}/{4}(|\Phi_+\rangle\langle\Phi_+|-|\Phi_-\rangle\langle\Phi_-|)$, yielding two dressed states $|\Phi_\pm\rangle\equiv\left(|\chi_2^M\rangle\pm|\chi_4^{M}\rangle\right)/\sqrt 2$ with energies $\pm \lambda=\pm{\sqrt M \bar{\Omega}_c}/{4}$. Then in the frame of  $\hat{h}'_1$ with a unitary transformation $\exp(i\hat{h}'_1t)$, Eq.~(\ref{S5}) becomes
\begin{equation}\label{Se2}
	\hat{\mathcal{H}}'_{\rm eff}= \sum_{M=0}^N\Big[\frac{\Omega_{t}}{2\sqrt2}|\chi_1^M\rangle\left(\langle \Phi_+|e^{-i\lambda t}+\langle \Phi_-|e^{i\lambda t}\right)\Big]+{\rm H.c.},
\end{equation}
for which the schematic diagram of transitions is shown in Fig.~\ref{f3}{(b)}. Here we choose $\bar{\Omega}_c/2\gg \Omega_{t}$, such that the transitions with $M\neq0$ are effectively suppressed, due to the faster oscillation at frequency $\lambda$. We then neglect the terms with $M\neq0$, so a final effective Hamiltonian is obtained
\begin{equation}\label{Se3}
	\hat{{H}}_{\rm e}=\Big(\bigotimes_{j=1}^N|D\rangle_j\langle D|\Big)\otimes{\hat H}_{t}.
\end{equation}
Only when all control qubits populate in $|D\rangle$ will the operation field on the target qubit work. Otherwise, the evolution of all the ($N+1$) qubits is frozen. It indicates that the controlled $(N+1)$-QHG can be achieved as long as a single-qubit holonomic gate $\hat U_{t}$ is operated on the target qubit.

\begin{figure*}\centering
	\includegraphics[width=0.8\linewidth]{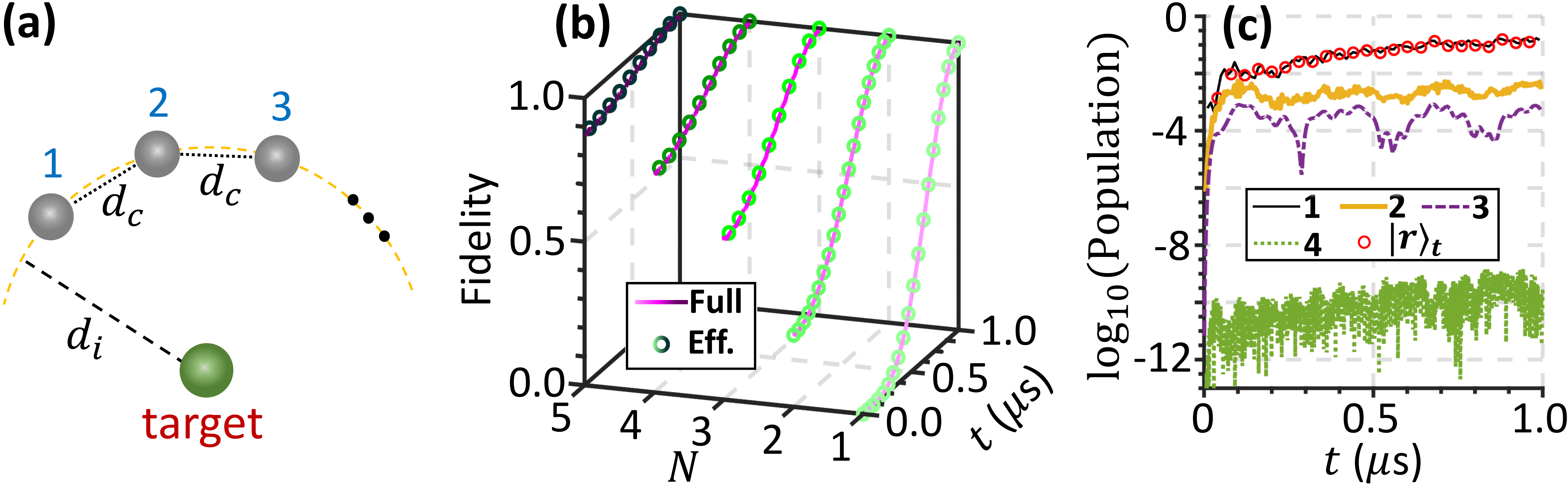}
	\caption{Rydberg atom setting and many-body dynamics. (a)~Spatial configuration of the Rydberg atom array. The target-control qubit distance is $d_i=3.8~\mu$m. The control atoms are distributed uniformly on a ring with the nearest-neighbor separation to be $d_c=2~\mu$m, and are driven homogeneously by the laser fields. (b)~Evolution of the C$_N$-NOT gate fidelity based on the full~(solid lines) and effective~(circles) Hamiltonian simulations, showing excellent agreement. (c)~Population dynamics of the collective states with different numbers of $|r\rangle$, when implementing the C$_3$-NOT gate with four atoms. The strong RRI prohibits the multiple Rydberg states from excitation. The red dotted line denotes the population of the target atom in state $|r\rangle_t$. Here $\bar{\Omega}_c=2\pi\times40~{\rm MHz}$ and $\Omega_t=2\pi\times 1~{\rm MHz}$. Other parameters are given in the text.}\label{f4}
\end{figure*}
\subsection{Multiqubit holonomic quantum gate}
Here we employ an established holonomic gate scheme on the target qubit~\cite{Sjoqvist2012,Xu2012,Abdumalikov2013,Feng2013,Zu2014,Arroyo2014,Liu2019,Hong2018,Sekiguchi2017,Zhou2017,Long2017,Xue2018,Zhang2019,Zhao2020,Yan2019}.
According to the Hamiltonian $\hat H_t$, one can design a suitable pulse scheme within duration $t\in[0,~T]$ such that a cyclic state transfer $|B\rangle_t\mapsto|A\rangle_t$ during $t\in[0,T/2]$ and then $|A\rangle_t\mapsto|B\rangle_t$ during $t\in(T/2,T]$ can be achieved without accumulating any dynamical phase. When we set the phases $\varphi_0=0$ and $\varphi_1=\pi+\phi$ for $t\in[0,~T/2]$, and switch them into $\varphi_0=\pi+\gamma$ and $\varphi_1=\gamma+\phi$ for $t\in(T/2,~T]$, an extra geometric phase $\gamma$ on $|B\rangle$ is obtained, for which the system experiences a single-loop geometric evolution, as shown on the Bloch sphere in the inset box of Fig.~\ref{f2}. Switching the phase of a drive field has been realized experimentally when implementing non-adiabatic non-Abelian~\cite{Xue2018}, shortcut-to-adiabatic non-Abelian~\cite{Yan2019}, non-adiabatic Abelian~\cite{Xue2020}, and shortcut-to-adiabatic Abelian~\cite{Kleibler2018} geometric gates. Finally, the state transformation of the target qubit can be governed by an evolution operator $|D\rangle_t\langle D|+e^{i\gamma}|B\rangle_t\langle B|$, which is expressed by the computational states $\{|0\rangle_t,|1\rangle_t\}$ as
\begin{eqnarray}
	\hat U_t&=&\left[
	\begin{array}{cc}
		\cos^2\frac{\theta}2+e^{i\gamma}\sin^2\frac{\theta}2 &~\frac12e^{-i\phi_1}(e^{i\gamma} -1)\sin\frac{\theta}2\\
		\frac12e^{i\phi_1}(e^{i\gamma} -1)\sin\frac{\theta}2               &~e^{i\gamma}\cos^2\frac{\theta}2+\sin^2\frac{\theta}2\\
	\end{array}
	\right]\nonumber\\
	&=&\exp(i\frac{\gamma}2)\exp(-i\frac{\gamma}2\boldsymbol{n}\cdot\boldsymbol{\sigma}),
\end{eqnarray}
where $\boldsymbol{n}=(\sin\theta\cos\phi,~\sin\theta\sin\phi,~\cos\theta)$ is a three-dimensional unit vector and $\boldsymbol{\sigma}=(\hat\sigma^x,~\hat\sigma^y,~\hat\sigma^z)$ are the standard Pauli operators.
With different choices of \{$\theta$, $\phi$, $\gamma$\}, an arbitrary single-qubit rotation $\hat U_t$ on the target qubit can be achieved, as well as permitting to flexibly implement a universal set of holonomic gates~($N=0,1$) and also multiqubit holonomic gates~($N\geq 2$).

\section{Realization with a Rydberg atom array}\label{sec3}
Due to the large polarizability, Rydberg atoms experience strong and long-range van der Waals interaction $V_{jk}=C_6/R_{jk}^6$~(corresponding to the Ising interaction in Hamiltonian $\hat{H}$) with $C_6$ being the dispersion coefficient and $R_{jk}$ the inter-atom distance~\cite{Jaksch2000,Saffman2010,Labuhn16,Browaeys2020,Bluvstein2021}. We will demonstrate that the strong Rydberg-Rydberg interactions~(RRI), precise control of atom positions with optical tweezer arrays~\cite{Levine2018,Omran2019,Graham2019,Levine2019,Jo2020} and of laser pulses allow to achieve high-fidelity multiqubit gates.

\subsection{Effective dynamics}
Our setting is a 2D atom array~\cite{Nogrette2014} where the control atoms sit on a ring and the target atom in the center, which ensures equivalent control-target interatom couplings, as depicted in Fig.~\ref{f4}(a). A pair of hyperfine ground states of atoms are used to encode qubit states $|0\rangle$ and $|1\rangle$, and a high-lying Rydberg state $|r\rangle$ corresponds to the auxiliary state, i.e., $|A\rangle=|r\rangle$. We specify $|B\rangle_j=|0\rangle_j$ and $|D\rangle_j=|1\rangle_j$ for all control atoms so that only $|0\rangle$ of control atoms is transited to the Rydberg state, while for the target atom we follow the general definitions. This yields the many-atom Hamiltonian
\begin{equation}\label{e3}
	\hat{H}_{I} = {\hat H}_{c} +  {\hat H}_{t}+ {\hat H}_{i},
\end{equation}
where ${\hat H}_{c}=\sum_{j=1}^N\frac{\Omega_{c}}2|0\rangle_j\langle r|+{\rm H.c.}$ and ${\hat H}_{t}=\sum_{s=0,1}\frac{\Omega_s}2e^{i \varphi_s}|s\rangle_{t}\langle r|+{\rm H.c.}=\frac{\Omega_t}2|B\rangle\langle r|+{\rm H.c.}$ with $\Omega_t=\sqrt{\Omega_0^2+\Omega_1^2}$. The space dependent RRI is described by  ${\hat H}_{i}=\sum_{j>k=1}^N(V_{jk}|rr\rangle_{jk}\langle rr|+V_{j{t}}|rr\rangle_{j{t}}\langle rr|)$ with $V_{jk}$ being the coupling strength between the $j$-th and $k$-th atoms~\cite{Browaeys2020,Bluvstein2021,Basak2018,Lukin2001}. Although the interaction depends on distances between the control qubits, we will show high-fidelity gates can be still achieved.

	In the Rydberg atom setting, one can choose parameters~(as given bellow) such that the interaction between any pair of the control quibts is relatively strong. This allows that maximally one control qubit can be excited to the Rydberg state. This corresponds to the desired situation with $K=M$ in the collective state $|\psi^K\rangle$.  Working in this restricted Hilbert space, we can obtain the approximation Hamiltonian, with the same form as Eq.~(\ref{e1})
\begin{eqnarray*}
	\hat{H}_{I}&\approx & \sum_{M=0}^N\Big[\frac{\sqrt{M} \Omega_{c}}{2}\Big({\hat S}_{31}^{(M)} e^{it\Delta_{31}^{(M)}}+{\hat S}_{42}^{(M)} e^{it\Delta_{42}^{(M)}}\Big)\nonumber\\
	&&+\frac{\Omega_{t}}{2}\Big({\hat S}_{21}^{(M)}e^{it\Delta_{21}^{{(M)}}}+{\hat S}_{43}^{(K)} e^{it\Delta_{43}^{(M)}}\Big)\Big]+{\rm H.c.}
\end{eqnarray*}
Then following similar derivation to obtain the effective Hamiltonian Eq.~(\ref{Se3}), Rabi frequencies
$\bar{\Omega}_c$ and $\Omega_{t}$ are chosen appropriately with $\omega=V_{j{t}}\gg\sqrt N|\bar{\Omega}_c|/4$ and $|\bar{\Omega}_c|\gg |\Omega_{t}|$, which guarantees that the underlying dynamics is governed by the effective Hamiltonian Eq.~(\ref{Se3}).

For the realization of the proposed many-body model with Rydberg atoms, to guarantee an important requirement that the target atom has an equal interaction to all of control atoms, the geometry of atom distribution is shown in Fig.~\ref{f4}(a), where a dozen of atoms can be loaded on the ring of radius $d_i=3.8~\mu$m and the interatomic separation $d_c=2~\mu$m. In addition to the 2D atomic array that can be readily achieved in experiment~\cite{Nogrette2014,Labuhn16}, a defect-free 3D array with the control atoms are distributed on a spherical surface may be also accessible~\cite{Barredo2018,Kumar2018}, based on which the available number of control atoms can be greatly increased. We remark here that, according to recent experiments, the neutral atom array with intersite spacing near~\cite{Labuhn16,Brown2019} or even less than~\cite{Glicenstein2021} $2~\mu$m is possible.

We emphasize that it is a ring that the control atoms are distributed on, so it is difficult to drive them homogeneously by a same field without individual addressing. Single-site addressability is needed, and the control atoms can be driven by $N$ independent fields with identical drive parameters, assisted by the devices of laser beam splitter. There is no obstacle to realize such single-site addressing and driving for Rydberg atoms in a 2D array by using drive lasers with waist $\leq1~\mu$m~\cite{Labuhn16,Bernien2017,Levine2018,Graham2019,Omran2019}~(see also the review article~\cite{Browaeys2020}).

\subsection{Multiqubit NHQC gates}
Specifically, we choose hyperfine ground states $|0(1)\rangle\equiv|5S_{1/2}, F=1(2), m_{F}=0\rangle$~\cite{Levine2019}, and a Rydberg state $|r\rangle\equiv|70S_{1/2}\rangle$ with $C_6/2\pi=858.4~{\rm GHz}\cdot\mu{\rm m}^6$ of $^{87}$Rb atoms~\cite{Omran2019,Walker2008,Levine2018,Levine2019,Bernien2017}. The transition $|0\rangle(|1\rangle)\leftrightarrow|r\rangle$ is driven through a two-photon process~(see Appendix~\ref{Appendix_B} for details), as demonstrated in recent experiments~\cite{Omran2019,Barredo2015,Levine2018,Levine2019,Jo2020,Bernien2017}. The control-target separation $d_i=3.8~\mu{\rm m}$ results to $V_{j{t}}=2\pi\times285.1$~MHz.  Rabi frequencies
$\bar{\Omega}_c$ and $\Omega_{t}$ are chosen appropriately with $\omega=V_{j{t}}\gg\sqrt N|\bar{\Omega}_c|/4$ and $|\bar{\Omega}_c|\gg |\Omega_{t}|$. This choice guarantees that the underlying dynamics is governed by the effective Hamiltonian~(\ref{Se3}). 

We illustrate that many-atom dynamics of the system indeed can be captured by the effective Hamiltonian~(\ref{Se3}). Such benchmark is carried out by implementing a C$_N$-NOT gate~\cite{Isenhower11,Khazali2020}. The figure of merit is the state overlap fidelity $F=|\langle \Psi_i|\Psi(t)\rangle|^2$, where $|\Psi_i\rangle$ is an ideal state after a C$_N$-NOT gate on the initial state $|\Psi_0\rangle=\bigotimes_j^N(|0\rangle_j-|1\rangle_j)\otimes(|0\rangle_{t}-|1\rangle_{t})/{\sqrt2}^{N+1}$, while the actual state $|\Psi(t)\rangle$ is obtained by solving the Schr\"odinger equation numerically. An important result is that the full Hamiltonian $\hat H_{I}$ and the effective Hamiltonian $\hat H_{e}$ produce nearly identical fidelity, as shown in Fig.~\ref{f4}(b). The excellent agreement results primarily from the fact that double or more excitations of the Rydberg state are strongly suppressed, which is verified in Fig.~\ref{f4}(c), validating the approximations used in deriving the effective Hamiltonian.

\begin{figure}\centering
	\includegraphics[width=\linewidth]{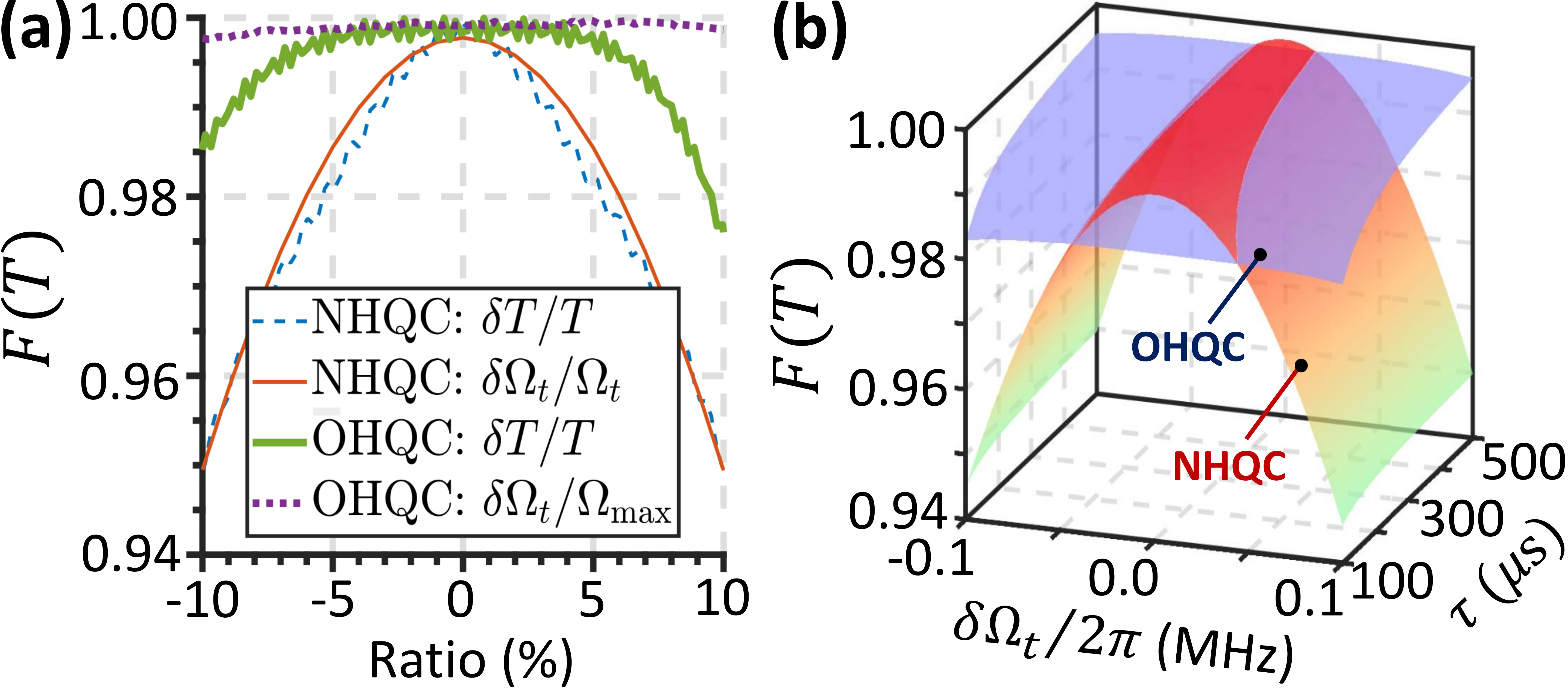}
	\caption{(a)~CNOT gate fidelity versus relative errors in $T$ and $\Omega_t$. The fidelity is improved significantly with OHQC. (b)~CNOT gate fidelity when varying  Rydberg  lifetime $\tau$ as well as $\delta\Omega_t$. OHQC is robust and achieves higher fidelities when $|\delta\Omega_t|/2\pi>30~{\rm kHz}$. The NHQC has high fidelities when $\delta \Omega_t$ is small.}\label{f5}
\end{figure}
Note that the physical regime is different from either the Rydberg blockade or antiblockade. In our protocol the first sideband of the modulation field offsets the control-target RRI, where $\omega=V_{j{t}}$ is similar to the Rydberg antiblockade condition~\cite{Su2017,Bai2020,WLi2013,Chen2018,Xing2020}. For a normal process of Rydberg antiblockade, a doubly-excited state can be achieved from $|\chi_1^M\rangle$ to $|\chi_4^M\rangle$, where the latter includes two Rydberg states. In our gate, however, only the target atom can be excited (i.e., the Rydberg blockade), as the strong control field prevents the excitation of doubly-excited Rydberg states. This feature is particularly beneficial to maintaining high gate fidelity, as state loss and motional dephasing due to Rydberg excitation are intrinsically mitigated. In contrast, other multiqubit gate schemes either encode the logic states in excited states~\cite{Rasmussen20}, or allow multiple control qubits in the excited-state manifold through, e.g., the three-step, adiabatic process~\cite{Khazali2020}, where the qubits could suffer stronger decay from the electronically excited states.

\section{Robust multiqubit holonomic gates}\label{sec4}
\subsection{Gate pulse engineering}
The NHQC gates are usually sensitive to gate lasers when the laser profile has a simple shape~\cite{Zheng2016,Liu2019,Liu2021}. To illustrate this dependence, we consider a rectangular pulse with small variation $\delta T$ ($\delta \Omega_t$) to the gate time $T$ (Rabi frequency $\Omega_t$). As depicted in Fig.~\ref{f5}(a), the infidelity of the CNOT gate of NHQC grows rapidly with increasing errors in $T$ or $\Omega_t$, showing the relatively large sensitivity of NHQC to the systematic errors.

It turns out that the fidelity as well as error tolerance of $(N+1)$-QHG can be improved through pulse engineering, permitting to carry out optimized holonomic quantum computation (OHQC)~\cite{Liu2019,Liu2021}. Here we achieve OHQC by employing a time-dependent Rabi frequency $\Omega_t(t)$ and detuning $\Delta(t)$ in the transition $|B\rangle_{t}\leftrightarrow|A\rangle_{t}$.
Due to the detuning, Hamiltonian~(\ref{e3}) now becomes $\hat H_I+\frac{\Delta(t)}2\hat Z_t$ with $\hat Z_t=|r\rangle_t\langle r|-|B\rangle_t\langle B|$. This results to an effective Hamiltonian, $\hat{H}'_{I}=(\bigotimes_j^N|1\rangle_j\langle1|)\otimes[\Omega_t(t)\hat X_t+\Delta(t)\hat Z_t]/2$, provided $\max\{|\Delta(t)|,|\Omega_t(t)|\}\ll\bar{\Omega}_c/2$. The pulse engineering is based on the optimal control technique~\cite{Ruschhaupt2012,Daems2013}, such that the time-dependent operation field is shaped elaborately for minimizing the systematic error sensitivity of OHQC gates~(see Appendix~\ref{Appendix_C} for details)
\begin{eqnarray}\label{e4}
	\Omega_{t}(t)&=&\dot{\alpha}\sqrt{1+\lambda^2\sin^2\alpha},\nonumber\\
	\Delta(t)&=&-\lambda\dot{\alpha}\cos\alpha-\frac{\dot{\lambda}\sin\alpha+\lambda\dot{\alpha}\cos\alpha}{1+\lambda^2\sin^2\alpha}.
\end{eqnarray}
$\alpha=12\pi t'^2/T^2-16\pi t'^3/T^3$ is a piecewise function  with $t'=t$ for $t\in[0,~T/2]$ and $t'=t-T/2$ for $t\in(T/2,~T]$, and $\lambda=2+2a_1\cos(2\alpha)+4a_2\cos(4\alpha)$. The optimal parameters are $a_1=0.28$ and $a_2 = -0.12$, corresponding to the gate time $\max\{\Omega_{t}(t)T\}/2\pi=2.85$. Piecewise phases are $\varphi_0=0$ and $\varphi_1=\pi+\phi$ during $t\in[0,~T/2]$ while $\varphi_0=\pi+\gamma$ and $\varphi_1=\gamma+\phi$ during $t\in(T/2,~T]$. This set of parameters not only results in geometric evolution shown on the Bloch sphere in Fig.~\ref{f2}, but also improves the gate tolerance to systematic errors. An example with $\Omega_{\rm max}\equiv\max\{|\Omega_t(t)|\}=2\pi\times1~{\rm MHz}$ is shown in Fig.~\ref{f5}(a). One sees that the pulse engineering leads to a robust and high-fidelity CNOT gate that is hardly affected by the fluctuation of Rabi frequency when $|\delta\Omega_t/\Omega_{\rm max}|\le0.1$.

\subsection{Optimized $(N+1)$-QHG}
We now thoroughly examine the robustness of OHQC where fluctuations of parameters relevant to current experiments will be considered. First, we show OHQC is largely immune to finite Rydberg lifetime. To take into account of spontaneous decay in the Rydberg state, a many-atom master equation is solved~(see Appendix~\ref{Appendix_D} for details), in which the qubit dephasing (rate $\gamma_{\phi}/2\pi=1~{\rm kHz}$) is included additionally. Through comparing the fidelity of a CNOT gate of OHQC with that of NHQC in Fig.~\ref{f5}(b), it is found that the gate fidelity of OHQC is robust, and higher when fluctuations of the Rabi frequency ranges within $|\delta\Omega_t|/2\pi>30~{\rm kHz}$. With typical experimental conditions, our extensive calculations furthermore show that fidelities for two- and three-qubit controlled-$\hat{U}_{t}(\theta,\phi,\gamma)$ gates of OHQC can be better than $0.995$ by identifying the average fidelity, as discussed in Appendix~\ref{Appendix_D}.

\begin{figure}\centering
	\includegraphics[width=\linewidth]{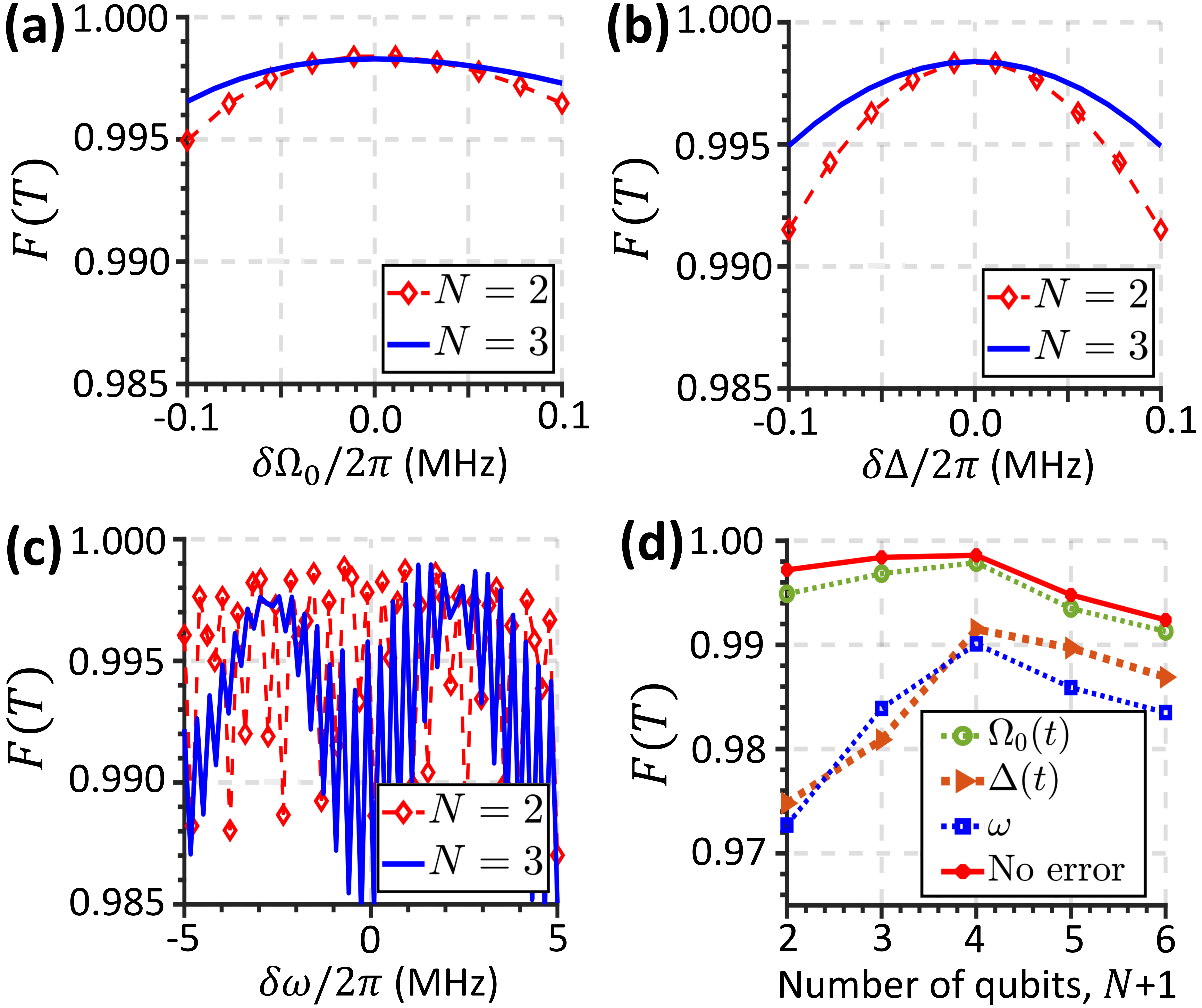}
	\caption{Error tolerance of the multiqubit holonomic gate. Fidelities of the C$_N$-NOT gate with $N=2$ and $N=3$ when  introducing errors in (c) $\Omega_0(t)$, (d) $\Delta(t)$ , and (e) $\omega$, respectively. (f)~C$_N$-NOT gate fidelity for different $N$. The solid and dotted line shows the fidelity without errors and when the parameter suffers $5\%$ relative errors, correspondingly.}\label{f6}
\end{figure}
Next, we demonstrate that fluctuations of the laser parameters only affect the gate fidelity marginally. For $N=2$ and $N=3$, we identify the tolerance of C$_{N}$-NOT gates to the systemic error in $\Omega_0$ in Fig.~\ref{f6}(a), and fidelities are always larger than 0.994 when $|\delta\Omega_0|/2\pi\le0.1~{\rm MHz}$. Errors in $\Delta(t)$, however, need attention, which can alter the gate time and lead to unwanted dynamical phase. The latter can be eliminated through applying spin-echo~\cite{Hahn1950,Yan2019}, then from Fig.~\ref{f6}(b) we can find the fidelity is barely reduced. On the other hand, the gate fidelity oscillates with varying $\delta\omega$ in Fig.~\ref{f6}(c), as non-zero $\delta\omega$ gives off-resonant coupling. However the resulting fidelity is rather high, around and above 0.99 when $|\delta\omega|/2\pi<5$~MHz.

We emphasize that the driving strength $\Omega_t(t)$ in the effective Hamiltonian~(\ref{Se3}) is not degraded, compared to the original one. In contrast, the effective driving strengths in other multiqubit gate schemes decrease with increasing number of qubits~\cite{Kang2019,Xing2020}. In fact the fidelity increases with $N$ in certain parameter regions, as shown in Figs.~\ref{f6}(a) and (b). Although it becomes difficult to achieve $V_{j{t}}\gg\sqrt N\bar{\Omega}_t$ when $N$ is too large, high fidelities are still obtained with moderate number of qubits, up to $N=5$ according to Figs.~\ref{f4}(b) and \ref{f6}(d).

\begin{figure}[t]\centering
	\includegraphics[width=\linewidth]{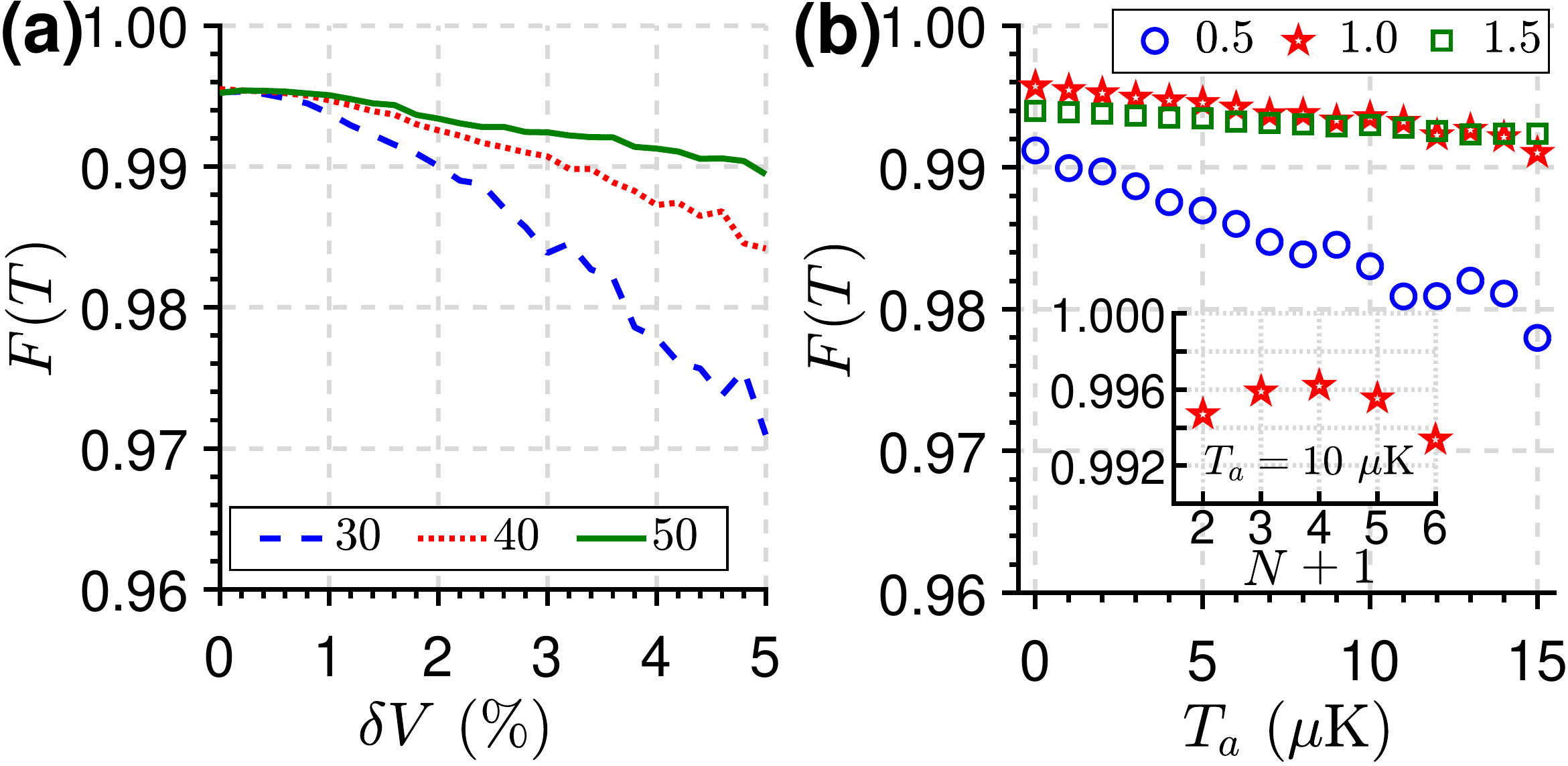}
	\caption{Gate robustness against the RRI fluctuation and finite temperature. (a)~CNOT gate fidelity versus relative errors of the RRI. The gate becomes robust with increasing $\bar{\Omega}_c/2\pi$ (labeled in the figure, unit MHz). (b)~CNOT gate fidelity decreases gradually when increasing atomic temperature $T_a$. Larger $\Omega_{\rm max}/2\pi$~(labeled in the figure, unit MHz) can strengthen the robustness. The inset shows the C$_N$-NOT gate fidelity with $\Omega_{\rm max}/2\pi=1$~MHz and $T_a=10~{\rm \mu K}$, which remains high when $N=5$~\cite{Levine2018,Levine2019}. Each point is an average of 201 realizations.}\label{f7}
\end{figure}
Finally, the effect of motional dephasing, due to imperfect positions of atoms in the trap array or thermal motions, can be suppressed, too. The imperfect placement of atoms causes errors to the RRI. To be specific, we assume the RRI fluctuates randomly in $[-\delta V,\delta V]$. Though the infidelity can be large with increasing $\delta V$ in Fig.~\ref{f7}(a), we can reduce its impact by increasing $\bar{\Omega}_c$, ensuring the system stays in the regime described by Hamiltonian~(\ref{Se3}). Thermal motions lead to unwanted detunings of pumping lasers seen by the atoms~\cite{Levine2018,Saffman2016,Leseleuc2018}. Our numerical simulations show that the thermal motions can be almost suppressed by increasing $\Omega_{\rm max}$, when jointly utilizing spin echo~\cite{Hahn1950,Levine2018}, as shown in Fig.~\ref{f7}(b). The thermal motion has negligible effects on the fidelity of the multiqubit gate even for $N=5$ at $T_a=10~\mu$K [inset Fig.~\ref{f7}(b)]. The high fidelity results from the fact that the effective dynamics is largely captured by Eq.~(\ref{Se3}).

\section{Conclusion}\label{sec5}
We have presented a robust scheme for realizing $(N+1)$-qubit holonomic gates. Its advantage is that errors due to electronic excitation of the control qubits are mitigated, enabled by the strong inter-spin interaction. Using Rydberg atom arrays, we have revealed the robustness of the multiqubit holonomic gate against errors in the laser parameters, variations of interatomic interaction, and  thermal fluctuations.
Besides neutral Rydberg atoms, trapped Rydberg ions~\cite{Zhang2020} are another candidate to realize the multiqubit gates due to the strong RRI. Moreover, the $(N+1)$-qubit holonomic gates can be implemented with superconducting circuits~\cite{SM1}. The very long-range coupling between superconducting qubits suggests that they are potentially less sensitive to the spatial fluctuation. Our study opens a new route to achieve robust and error-tolerant multiqubit holonomic gates with strongly interacting Rydberg atoms and superconducting circuits, which might find applications in scalable quantum computation and simulation of many-body models.

\section*{Acknowledgements}
		We acknowledge supports from National Natural Science Foundation of China (NSFC) (11675046, 11804308, 21973023), Program for Innovation Research of Science in Harbin Institute of Technology (A201412), Postdoctoral Scientific Research Developmental Fund of Heilongjiang Province (LBH-Q15060), and Natural Science Foundation of Henan Province Under Grant No.~202300410481. W. L. acknowledges support from the EPSRC through Grant No.~EP/R04340X/1 via the QuantERA project “ERyQSenS”, the UKIERI-UGC Thematic Partnership (IND/CONT/G/16-17/73), and the Royal Society through the International Exchanges Cost Share award No. IEC$\backslash$NSFC$\backslash$181078.
		
\appendix
\section{Full Hamiltonian in the frame of the Ising interaction, Eq.~(\ref{e1})}\label{Appendix_A}
	Despite the decoupled state $|D\rangle$, qubits considered in our work can be regarded as spin-$\frac12$ particles, where $|A\rangle$ and $|B\rangle$ denote positive and negative spin states, respectively. We introduce spin operators $\hat S^{(z)}_j\equiv(|A\rangle_j\langle A|-|B\rangle_j\langle B|)/2$, $\hat{S}_j^{+}=|A\rangle\langle B|$ and $\hat{S}_j^{-}=|A\rangle\langle B|$ for the $j$-th qubit, as well as the number operator $\hat n_j=\frac12+S_j^{(z)}$. When $M$ control qubits populate in \{$|A\rangle$, $|B\rangle$\} while other $(N-M)$ ones in $|D\rangle$, we define collective spin operators $\hat J^{(z)}_c\equiv\sum_{j}^M\hat S_j^{(z)}=\sum_{j}^M\hat n_j-\frac M2$ and $\hat{J}_c^{\pm}=\sum_j^M\hat{S}_j^{\pm}$. The collective eigenstates of the $M$ spin-$\frac12$ particles are expressed using the Dicke state, $|j,m\rangle$ with $j=M/2$, and $m=-j,(-j+1),\cdots,(j-1),j$, satisfying $\hat J^{(z)}_c|j,m\rangle=m|j,m\rangle$, and  $\hat J_c^{\pm}|j,m\rangle=\sqrt{j(j+1)-m(m\pm1)}|j,m\pm1\rangle$.

Now one can derive the Hamiltonian using the collective Dicke state. First, we divide the Ising interactions among $(M+1)$ qubits~(including the $M$ control qubits and the target qubit) to two parts $\hat{H}_{\rm Ising}=\hat{H}_{\rm Ising}^{(c)}+\hat{H}_{\rm Ising}^{(t)}$ with the control-control interactions $\hat{H}_{\rm Ising}^{(c)}=\sum_{j>k=1}^MJ_c{\hat n}_j{\hat n}_k$ and the control-target interactions $\hat{H}_{\rm Ising}^{(t)}=\sum_{j}^MJ_t{\hat n}_j{\hat n}_{t}$. Using $\hat J^{(z)}_c\equiv\sum_{j}^M\hat S_j^{(z)}=\sum_{j}^M\hat n_j-\frac M2$, we can rewrite the control-control interactions as
\begin{eqnarray}
	\hat{H}_{\rm Ising}^{(c)}&=&~\frac{J_c}2\sum_j^M\hat n_j\cdot\sum_{k\neq j}^M\hat n_k\nonumber\\
	&=&~\frac{J_c}2\sum_j^M\hat n_j\cdot\sum_{k}^M(\hat n_k-\hat n_k\delta_{jk})\nonumber\\
	&=&~\frac{J_c}2[\hat J^{(z)}_c+\frac{M}2][\hat J^{(z)}_c+\frac{M}2-1].
\end{eqnarray}
Similarly, then we can obtain $\hat{H}_{\rm Ising}^{(t)}=J_t[\frac12+\hat S_t^{(z)}][\frac M2+\hat J^{(z)}_c]$ with $\hat S_t^{(z)}$ being the spin operator of the target qubit, so $\hat{H}_{\rm Ising}$ is exactly described by the collective spin operator $\hat{J}_c^{(z)}$ and $\hat{S}_t^{(z)}$
\begin{eqnarray}
	\hat{H}_{\rm Ising} &=& \frac{J_c}2[\hat J^{(z)}_c+\frac{M}2][\hat J^{(z)}_c+\frac{M}2-1]\nonumber\\
	&&+J_t[\frac12+\hat S_t^{(z)}][\frac M2+\hat J^{(z)}_c].
\end{eqnarray}

Using the Dicke states, we re-express the control qubit state $|\psi^K\rangle\equiv|D\rangle^{N-M}|B\rangle^K|A\rangle^{M-K}=|\frac M2,m=\frac M2-K\rangle_c$ and the target qubit states, $|A\rangle_t=|\frac12,\frac12\rangle_t$ and $|B\rangle_t=|\frac12,-\frac12\rangle_t$. Here the value of $m$ is calculated according to the numbers of $|A\rangle$ and $|B\rangle$ in $|\psi^K\rangle$, that is, $m=(M-K)\times\frac12-K\times\frac12$. Then the composite basis of the collective control and target qubits denoted with $|\chi_1^K\rangle\equiv|\psi^K\rangle|B\rangle_{t}$, $|\chi_2^K\rangle\equiv|\psi^K\rangle|A\rangle_{t}$, $|\chi_3^K\rangle\equiv|\psi^{K-1}\rangle|B\rangle_{t}$, and $|\chi_4^K\rangle\equiv|\psi^{K-1}\rangle|A\rangle_{t}$, where there are $N_A=M-K$ ($N_A=M-K+1$) control qubits in state $|A\rangle$ for $|\chi_{1,2}^K\rangle$ ($|\chi_{3,4}^K\rangle$), can be re-expressed, respectively, by
\begin{eqnarray}
	&&|\chi_1^K\rangle=|\frac M2,\frac M2-K\rangle_c\otimes|\frac12,-\frac12\rangle_t,\nonumber\\
	&&|\chi_2^K\rangle=|\frac M2,\frac M2-K\rangle_c\otimes|\frac12,\frac12\rangle_t,\nonumber\\
	&&|\chi_3^K\rangle=|\frac M2,\frac M2-K+1\rangle_c\otimes|\frac12,-\frac12\rangle_t,\nonumber\\
	&&|\chi_4^K\rangle=|\frac M2,\frac M2-K+1\rangle_c\otimes|\frac12,\frac12\rangle_t,
\end{eqnarray}
which are eigenstates of $\hat{H}_{\rm Ising}$ with eigenvalues
\begin{eqnarray}\label{R4}
	&&E_1^{(K)}=\frac{J_c}2(M-K)(M-K-1),\nonumber\\
	&&E_2^{(K)}=\frac{J_c}2(M-K)(M-K-1)+J_t(M-K),\nonumber\\
	&&E_3^{(K)}=\frac{J_c}2(M-K+1)(M-K),\nonumber\\
	\cr&&E_4^{(K)}=\frac{J_c}2(M-K+1)(M-K)+J_t(M-K).\nonumber\\
\end{eqnarray}
The eigenvalues in Eq.~(\ref{R4}) can be unified as $E_p^{(K)}=C_{N_A}^2J_c +\xi N_AJ_t$ for state $|\chi_p^K\rangle$  with $\xi =0$ ($\xi=1$) for $p=1,3$ ($p=2,4$) and $C_{N_A}^2$ to be the binomial coefficient. Hence one can diagonalize the Ising interaction to be $\hat{H}_{\rm Ising}=\sum_{M=0}^N\sum_{K=0}^M\sum_{p=1}^4E_p^{(K)}|\chi_p^K\rangle\langle\chi_p^K|$.

Now we can calculate the collective coupling strength between $|\psi^K\rangle$ and $|\psi^{K-1}\rangle$~($1\leq K\leq M$)
\begin{eqnarray}\label{R5}
	\hat J_c^+|\psi^K\rangle&=&~\hat J_c^+|\frac M2,\frac M2-K\rangle_c\nonumber\\
	&=&~\sqrt{K(M-K+1)}~|\frac M2,\frac M2-K+1\rangle_c\nonumber\\
	&=&~\sqrt{K(M-K+1)}~|\psi^{K-1}\rangle.
\end{eqnarray}
Accordingly, there is a strengthened collective Rabi frequency $\sqrt{K'}\Omega_c$ with $K'={K(M-K+1)}$ for the coupling between the two Dicke states $|\psi^K\rangle$ and $|\psi^{K-1}\rangle$. Therefore, the Hamiltonian of two fields interacting with the $(N+1)$ qubits can be written as
\begin{eqnarray}\label{R6}
	\hat{H}_{fq}&=&\sum_{M=0}^N\sum_{K=0}^M\Big[\frac{\sqrt {K'} \Omega_{c}}{2}\Big(\hat S_{31}^{(K)} +\hat S_{42}^{(K)}\Big)\nonumber\\
	&&+\frac{\Omega_{t}}{2}\Big(\hat S_{21}^{(K)} +\hat S_{43}^{(K)}\Big)\Big]+{\rm H.c.},
\end{eqnarray}
where ${\hat S}_{pq}^{(K)}=|\chi_p^K\rangle\langle\chi_q^K|$ are collective transition operators. Finally we can see the total Hamiltonian, $\hat{H}_{\rm Ising}+\hat{H}_{fq}$, of the whole system as Eq.~(\ref{e1}).

\begin{figure}[htb]
	\includegraphics[width=0.6\linewidth]{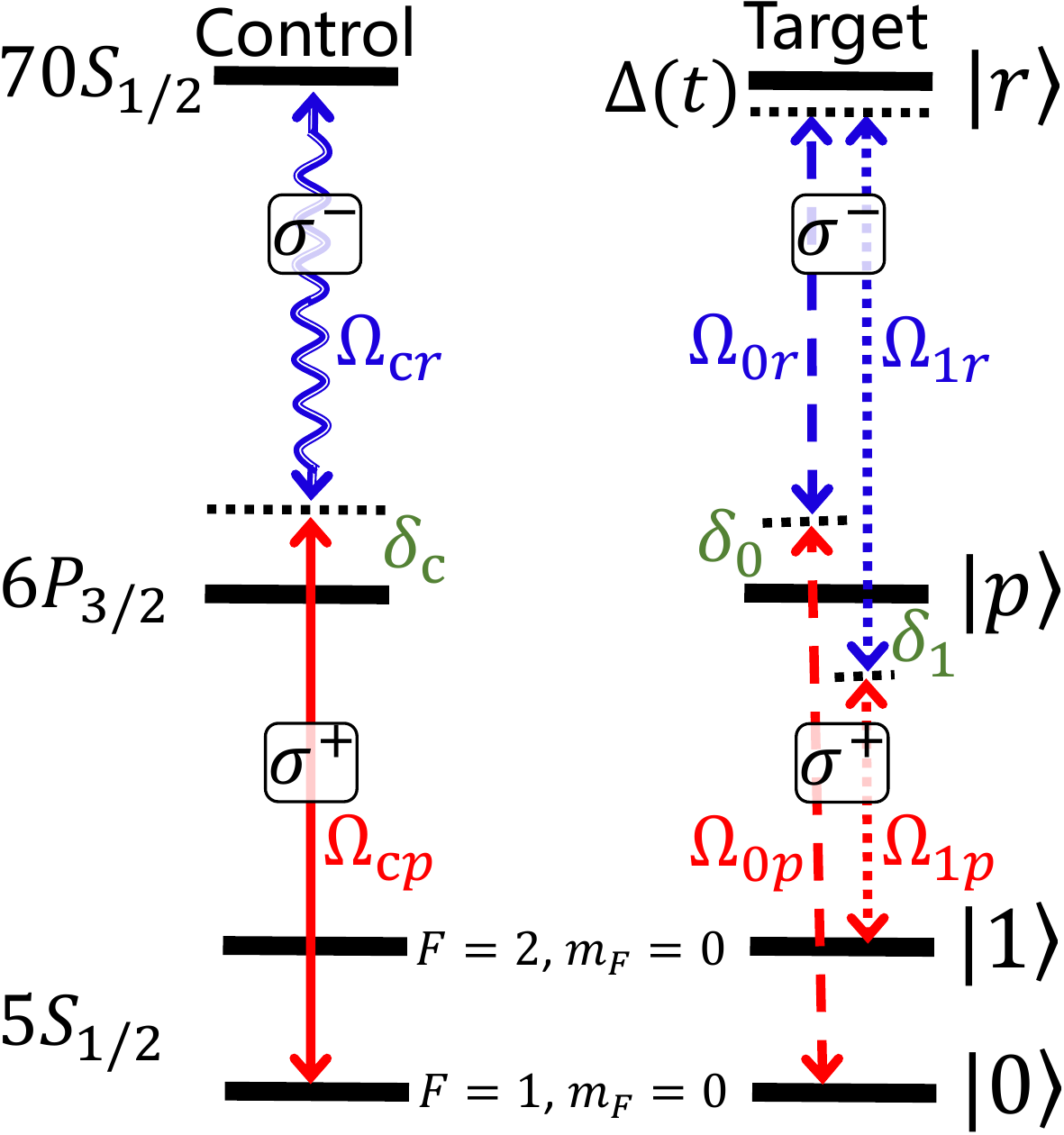}
	\caption{Schematic diagram for the two-photon Rydberg excitations in $^{87}$Rb atoms. $\sigma^{\pm}$ denote the polarization of lasers.}\label{f8}
\end{figure}
\section{Two-photon Rydberg pumping}\label{Appendix_B}
In order to implement holonomic gates with Rydberg atoms, the Rydberg pumping from a ground state to a Rydberg
state can be achieved by a two-photon process~\cite{Barredo2015,Bernien2017,Levine2018,Levine2019,Omran2019,Jo2020} in Rb atoms or a single-photon process~\cite{Graham2019} in Cs atoms. Here we consider related energy levels of $^{87}$Rb atoms $|0\rangle\equiv|5S_{1/2}, F = 1, m_{F}=0\rangle$, $|1\rangle\equiv|5S_{1/2}, F = 2, m_{F}=0\rangle$, and $|r\rangle\equiv|70S_{1/2}, m_{J}=-1/2\rangle$~\cite{Levine2019} with $C_6/2\pi=858.4~{\rm GHz}\cdot\mu {\rm m}^6$~\cite{Bernien2017}. Then assisted by an intermediate state $|p\rangle=|5p_{3/2}\rangle$, the atomic transition $|0(1)\rangle=|5S_{1/2}, F=1(2), m_{F}=1(2)\rangle\leftrightarrow|r\rangle=|70S_{1/2}\rangle$ can be achieved through a two-photon process, as shown in Fig.~\ref{f8}.

\begin{figure}[b]\centering
	\includegraphics[width=\linewidth]{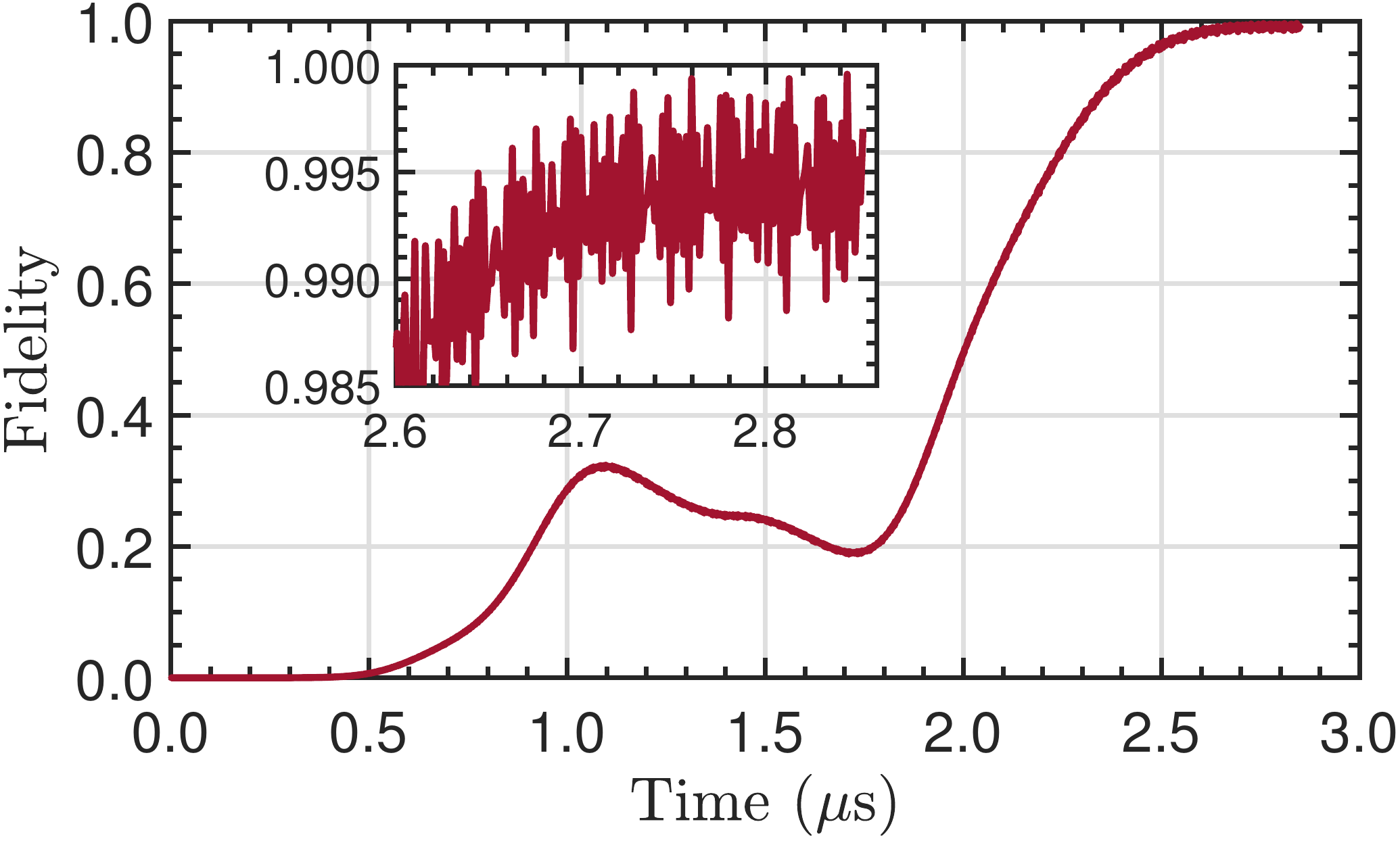}
	\caption{Fidelity of performing a CNOT gate  on an initial state $(|0\rangle_1-|1\rangle_1)\otimes(|0\rangle_{t}-|1\rangle_{t})/2$ based on the four-level atoms shown in Fig.~\ref{f8} for examining the validity of the two-photon processes by using the OHQC scheme. The inset shows the fidelity towards the end of the gate.}\label{f9}
\end{figure}
Each transition is achieved by two laser fields. For the control atoms, one field is imposed for the optical excitation $|0\rangle\leftrightarrow|p\rangle$ with Rabi frequency $\Omega_{{\rm c}p}$ and a blue detuning $\delta_{c}$, and the other for the Rydberg excitation $|p\rangle\leftrightarrow|r\rangle$ with Rabi frequency $\Omega_{{\rm c}r}$ and a red detuning $\delta_{c}$. The field on the control atoms for $|p\rangle\leftrightarrow|r\rangle$ is modulated in amplitude $\Omega_{cr}=\bar{\Omega}_{cr}\cos\omega t$. For the target atom, one field is imposed for the optical excitation $|0\rangle\leftrightarrow|p\rangle$~($|1\rangle\leftrightarrow|p\rangle$) with (time-dependent) Rabi frequency $\Omega_{{\rm 0}p}(t)$~[$\Omega_{{\rm 1}p}(t)$] and a blue~(red) detuning $\delta_{0}$~($\delta_1$), and the other for the Rydberg excitation $|p\rangle\leftrightarrow|r\rangle$ with Rabi frequency $\Omega_{{\rm 0}r}$~($\Omega_{{\rm 1}r}$) and a red~(blue) detuning $\delta_{\rm 0}+\Delta(t)$~[$\delta_1-\Delta(t)$]. The detunings for $|0\rangle\leftrightarrow|p\rangle$ and $|1\rangle\leftrightarrow|p\rangle$ are of opposite signs so as to avoid the effective coupling between two ground states. $\Delta(t)$ is a small time-dependent detuning for pulse engineering, while $\delta_{c}$, $\delta_{0}$, and $\delta_{1}$ are so large that the intermediate state $|p\rangle$ can be adiabatically eliminated and the Rydberg pumping from ground states is achieved.

The quantities used directly for holonomic gates are $\bar{\Omega}_c=\bar{\Omega}_{cr}\Omega_{{\rm c}p}/2\delta_{c}$, $\Omega_0(t)=\Omega_{0r}\Omega_{0p}(t)/2\delta_0$, and $\Omega_1(t)=\Omega_{1r}\Omega_{1p}(t)/2\delta_1$.
It should also be noted that in addition to the two-photon effective coupling from a ground state to the Rydberg state, there are Stark-shift terms $\Omega_{{\rm c}p}^2/4\delta_{{\rm c}}|0\rangle_{{\rm c}}\langle0|$, $\bar{\Omega}_{cr}^2\cos^2\omega t/4\delta_{{\rm c}}|r\rangle_{{\rm c}}\langle r|$, $\Omega_{0p}^2/4\delta_0|0\rangle_{{\rm t}}\langle0|$, $-\Omega_{1p}^2/4\delta_1|1\rangle_{{\rm t}}\langle1|$, and $(\Omega_{0r}^2/4\delta_0-\Omega_{1r}^2/4\delta_1)|r\rangle_{{\rm t}}\langle r|$. These unwanted energy shifts can be offset by imposing additional lasers to drive the off-resonant transitions between the related states and some auxiliary states so as to induce opposite Stark shifts~\cite{JLWu2022,JXHan2020}. Alternatively, these unwanted energy shifts can also be eliminated through phase corrections~\cite{Vepsaaineneaau5999}.

\begin{figure}[htp]
	\includegraphics[width=\linewidth]{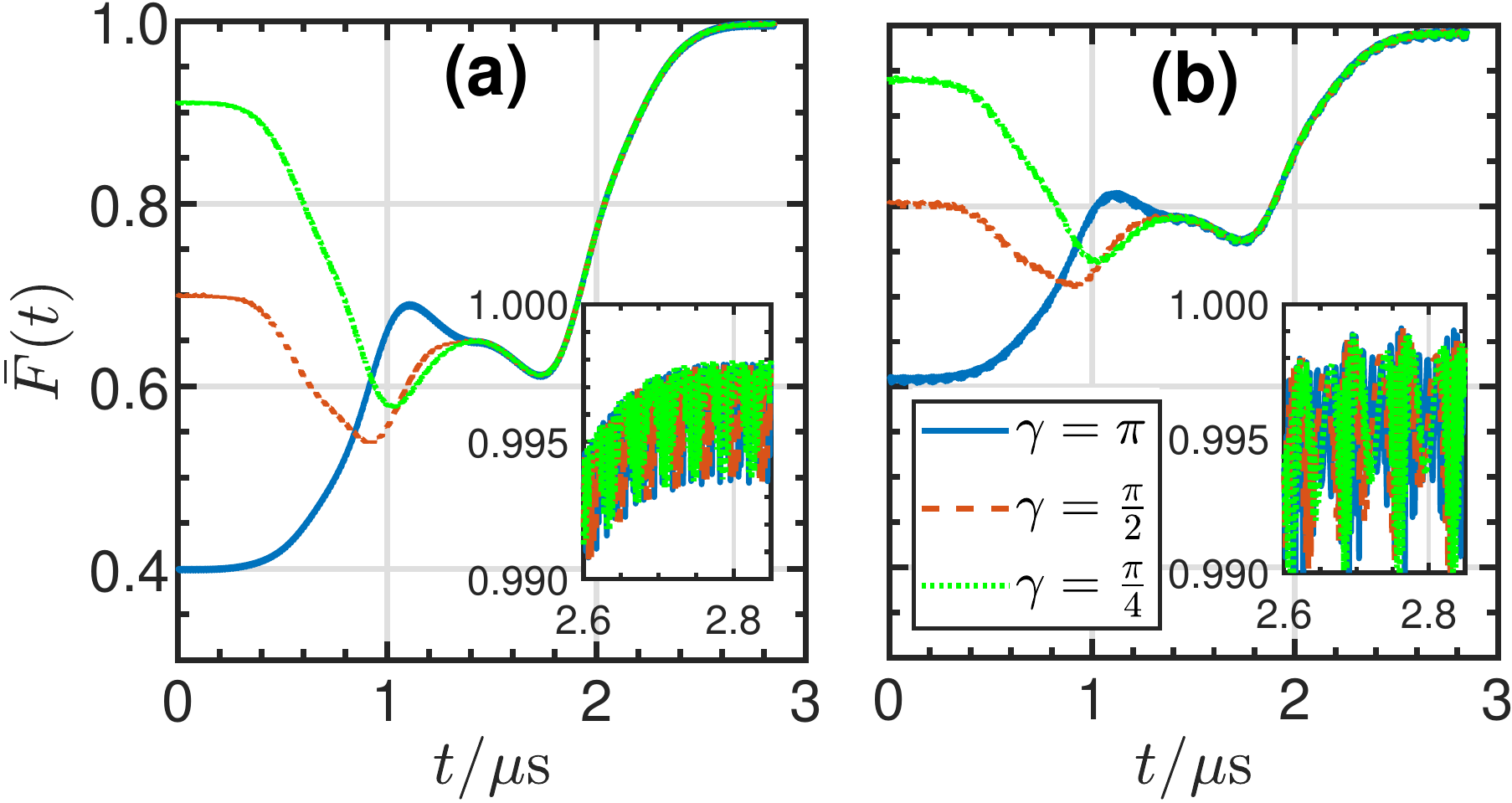}
	\caption{Trace-preserving-operator-based average fidelities of controlled $\hat{U}_{t}(\theta,\phi,\gamma)$ gates with different values of $\gamma$. {(a)}~Two-qubit gates. {(b)}~Three-qubit gates. $\tau=400~{\rm \mu s}$.}\label{f11}
\end{figure}

To test the validity of the two-photon processes discussed above, we use the four-level atoms shown in Fig.~\ref{f8} to perform a CNOT gate based on the OHQC scheme~(see Appendix~\ref{Appendix_C} for details of optimal pulses). We adopt parameters ${\Omega_{cp}=\bar\Omega}_{cr}=2\pi\times400~$MHz, $\delta_c=2\pi\times2$~GHz, $\Omega_{0r}=\max[\Omega_{0p}(t)]=\Omega_{1r}=\max[\Omega_{1p}(t)]=2\pi\times60/\sqrt{2}~$MHz, $\delta_0=\delta_1=2\pi\times1.8/\sqrt{2}$~GHz, and $V=\omega=2\pi\times285.1$~MHz. The fidelity evolution of the target state based on Schr\"odinger equation simulation is shown in Fig.~\ref{f9} for OHQC. The fidelity can reach  $0.995$, which is high, compared with results based on the effective three-level atoms used to illustrate gate performances in our work; see the contents related to the average gate fidelities shown in Fig.~\ref{f11} based on the effective three-level atoms, where the final average gate fidelities oscillates around 0.995. This shows that the two-photon Rydberg pumping can be safely used to conduct our gate scheme, even when a fast-oscillating
driving amplitude $\Omega_{cr}$ is applied in obtaining the effective Hamiltonian.

\begin{figure}[b]
	\includegraphics[width=\linewidth]{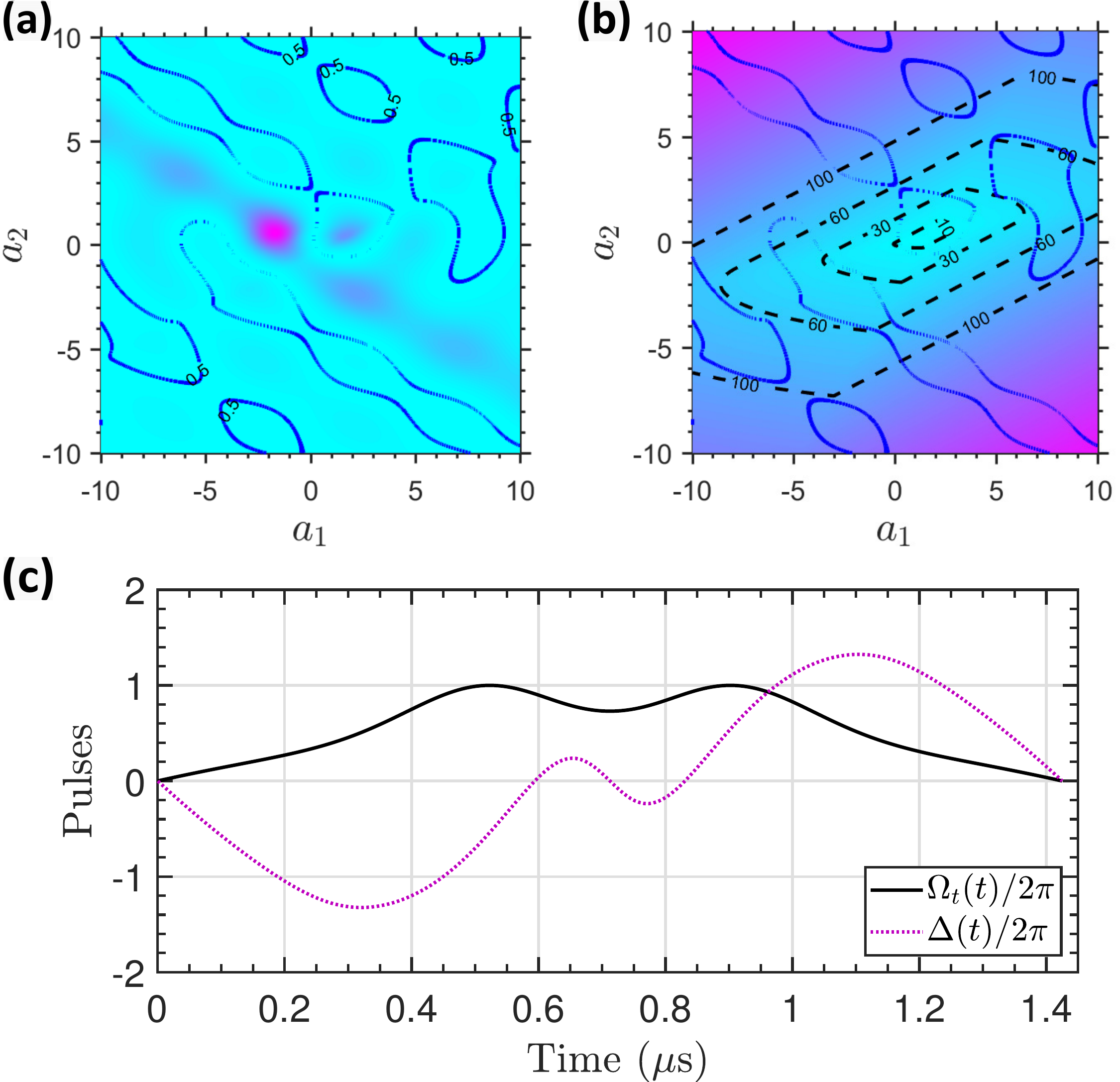}
	\caption{{(a)}~Error sensitivity $S$ versus $a_1$ and $a_2$. Blue curves are 0.5 contour lines of $S$. {(b)}~$T\max[\Omega_{t}(t)]/2$ versus $a_1$ and $a_2$. Black dashed curves are contour lines of $T\max[\Omega_{t}(t)]/2$ with different values. (c)~Pulse forms with $\max[\Omega_{t}(t)]/2\pi=1$~MHz, $a_1=0.28$ and $a_2 = -0.12$.}\label{f10}
\end{figure}
\section{Optimized pulse engineering}\label{Appendix_C}
Using the target-atom Hamiltonian $\hat{H}'_{t}=[\Omega_{t}(t)\hat X_{t}+\Delta(t)\hat Z_{t}]/2$ with $\hat X=|B\rangle\langle r|+|r\rangle\langle B|$ and $\hat Z=|r\rangle\langle r|-|B\rangle\langle B|$, we transfer the full population from $|B\rangle_{t}$ first to $|r\rangle_{t}$ during $t\in[0,T/2]$ and then back to $|B\rangle_{t}$ during $t\in(T/2,T]$, with $T$ being the gate time. In the following, we mainly discuss the population transfer from $|B\rangle_{t}$ to $|r\rangle_{t}$, because the inverse process is based on the same theory. During the state transfer from $|B\rangle_{t}$ to $|r\rangle_{t}$, a solution of the time-dependent Schr\"{o}dinger equation $i\partial|\psi\rangle/\partial t=\hat{H}'_{t}|\psi\rangle$ can be parametrized as a superposition $|\psi_0(t)\rangle=e^{-i\eta/2}\left[e^{i\beta/2}\cos(\alpha/2)|B\rangle_{t}+e^{-i\beta/2}\sin(\alpha/2)|r\rangle_{t}\right]$~\cite{Daems2013}, for which the time dependence symbol ``$(t)$" of $\alpha(t)$, $\beta(t)$, and $\eta(t)$ is omitted for simplicity. There is also an orthogonal solution $|\psi_1(t)\rangle=e^{i\eta/2}\left[ e^{i\beta/2}\sin(\alpha/2)|B\rangle_{t}-e^{-i\beta/2}\cos(\alpha/2)|r\rangle_{t}\right]$ such that $\langle\psi_1(t)|\psi_0(t)\rangle=0$. Inserting $|\psi_0(t)\rangle$ and $|\psi_1(t)\rangle$ into the Schr\"{o}dinger equation, $\Omega_{t}(t)$ and $\Delta(t)$ are related to $\alpha$ and $\beta$, as
\begin{equation}\label{e8}
\Omega_{t}(t)=\frac{\dot{\alpha}}{\sin\beta},\quad\Delta(t)=\dot{\beta}-\dot{\alpha}\cot\alpha\cot\beta.
\end{equation}
At the same time, the global phase $\eta=\int_0^t\dot{\alpha}(t')\cot\beta(t')/\sin\alpha(t') dt'$ can be obtained, and $\eta(T)=0$ is supposed to be satisfied so as to ensure a null dynamical phase. We assume that the evolution follows $|\psi_0(t)\rangle$, and thus the boundary conditions $\alpha(0)=0$ and $\alpha(T/2)=\pi$ are supposed to be satisfied for completing the state transfer $|B\rangle_{t}\mapsto|r\rangle_{t}$.

We introduce a systematic error $\delta{X}$ into an ideal parameter $X$, yielding an actual parameter $(X+\delta{X})$. Our goal is to design a pair of $\Omega_{t}(t)$ and $\Delta(t)$ such that the holonomic gates are insensitive to systematic errors in the gate time $T$ and the pulse amplitude $\Omega_{t}(t)$. For errors in the gate time, $\Omega_{t}(t)$ and $\Delta(t)$ can be designed to be softly turned on and off, so that a moderate surplus or deficiency in the pulse duration has little effect on the pulse area. When $\delta{\Omega_{t}}$ is taken into account, it leads to a perturbation-containing Hamiltonian $\hat{\mathcal{H}}_{t}=\hat{H}'_{t}+\hat{\mathcal{H}}_r$ with $\hat{\mathcal{H}}_r=\frac{\delta\Omega_{t}}2 \hat X_{t}$. Then using the perturbation theory and keeping the final state to the second order, we obtain a perturbed population of $|r\rangle_t$ at $t=T/2$, as 
$P_r(T/2)\simeq1-\left|\int_0^{T/2}dt\langle\psi_1(t)|\hat{\mathcal{H}}_r|\psi_0(t)\rangle\right|^2$. Then we define a quantity $S\equiv-\frac12\left.\frac{\partial^2P_r}{{\partial\delta{\Omega_{t}^2}}}\right|_{\delta{\Omega_{t}}=0}$ to measure the sensitivity of $P_r$ to the systematic error in $\Omega_{t}(t)$~\cite{Ruschhaupt2012}. Substituting the expressions of $|\psi_0(t)\rangle$ and $|\psi_1(t)\rangle$ into $P_r(T/2)$, the systematic error sensitivity can be calculated out
\begin{equation}\label{e9}
S=\frac14\left|\int_0^{T/2}dte^{-i\eta}(\cos\alpha\cos\beta+i\sin\beta)\right|^2.
\end{equation}
For rendering holonomic gates to hold strong tolerance to $\delta\Omega_t$, $S$ is supposed to be as small as possible, which needs pulse engineering with suitable forms of $\alpha$, $\beta$, and $\eta$.
According to $\alpha(0)=0$ and $\alpha(T/2)=\pi$, $\alpha$ can be designed with a polynomial ansatz as $\alpha=12\pi t^2/T^2-16\pi t^3/T^3$. Then we choose $\eta=2\alpha+a_1\sin(2\alpha)+a_2\sin(4\alpha)$ such that the dynamical phase is absent at $t=T/2$, and $a_1$ and $a_2$ are to be determined so as to minimize $S$.
The forms of $\alpha$ and $\eta$ give $\beta=\cos^{-1}(\lambda\sin\alpha/\sqrt{1+\lambda^2\sin^2\alpha})$ with $\lambda=2+2a_1\cos(2\alpha)+4a_2\cos(4\alpha)$, so the pulse form given in Eq.~(\ref{e4}) is obtained.

In order to achieve a small sensitivity to systematic errors and a short gate time, in Figs.~\ref{f10}{(a)} and {(b)} we plot numerically $S$ and $T\max[\Omega_{t}(t)]/2$, respectively, with varying $a_1$ and $a_2$. We find that in the region of $T\max[\Omega_{t}(t)]/2<30$ there exists a very small region of $S<0.5$, which indicates a trend that a small error sensitivity costs a longer gate time. However, there are still points guaranteeing $S<0.5$ and $T\max[\Omega_{t}(t)]/2<10$. For example, $a_1=0.28$ and $a_2 = -0.12$ give $S=0.3$ and $T\max[\Omega_{t}(t)]=17.9$. This gate time is three times longer than that~($2\pi$) of non-adiabatic holonomic gates.

With the trade-off between the robustness and the speed of implementing holonomic gates, we pick up $a_1=0.28$ and $a_2 = -0.12$ that can ensure a short gate time $\max[\Omega_{t}(t)]T/2\pi=2.85$ and a small systematic-error sensitivity $S=0.3$. Based on Eq.~(\ref{e4}), the pulse forms can be determined, as shown in Fig.~\ref{f10}(c), with which the state transfer $|B\rangle_{t}\mapsto|r\rangle_{t}$ can be achieved with an enhanced tolerance against the systematic error in $\Omega_{t}(t)$. An identical process can be performed again to transfer the population from $|r\rangle_{t}$ back to $|B\rangle_{t}$. In addition to optimal control for engineering pulses, other techniques may also be efficient to enhance gate robustness even with a constant Rabi frequency, for example Landau-Zener-St\"uckelberg interferometry~\cite{LZS2010,Wu2021}.

\section{Average fidelities of ($N+1$)-QHG}\label{Appendix_D}
The evolution of the many-body master equation can be simulated by numerically solving the Markov master equation based on the fourth-order variable step-size Runge-Kutta algorithm
\begin{eqnarray}\label{e11}
	{\dot{\hat{\rho}}}&=&-i[\hat{H}_{\rm I},\hat{\rho}]-\frac{1}{2}\sum^{N+1}_{j=1}\Big[\sum_{s=0}^2\Big(\hat{\mathcal{L}}_{js}^{\dag}\hat{\mathcal{L}}_{js}\hat{\rho}-2\hat{\mathcal{L}}_{js}\hat{\rho}\hat{\mathcal{L}}_{js}^{\dag}\nonumber\\
	\cr&& +\hat{\rho}\hat{\mathcal{L}}_{js}^{\dag}\hat{\mathcal{L}}_{js}\Big)+\Big(\hat{\mathcal{L}}_{\phi j}^{\dag}\hat{\mathcal{L}}_{\phi j}\hat{\rho}-2\hat{\mathcal{L}}_{\phi j}\hat{\rho}\hat{\mathcal{L}}_{\phi j}^{\dag} +\hat{\rho}\hat{\mathcal{L}}_{\phi j}^{\dag}\hat{\mathcal{L}}_{\phi j}\Big)\Big].\nonumber\\
\end{eqnarray}
$\hat{\mathcal{L}}_{js}\equiv\sqrt{\gamma_s}|s\rangle_{j}\langle r|$ and $\hat{\mathcal{L}}_{\phi j}\equiv\sqrt{\gamma_{\phi j}}(|r\rangle_{j}\langle r|-|B\rangle_{j}\langle B|)$ describe energy relaxation and dephasing of the $j$-th atom, respectively, with relaxation rate $\gamma_s$ from $|r\rangle$ to $|s\rangle$ and dephasing rate $\gamma_{\phi j}$ for coherence of $|r\rangle$ and $|B\rangle$. For the energy relaxation, an additional ground state $|2\rangle$ is introduced to denote those Zeeman magnetic sublevels out of the computational states $|0\rangle$ and $|1\rangle$. For convenience, we assume that energy relaxation rates from a Rydberg state of $^{87}$Rb atoms into the eight Zeeman ground states are identical, so $\gamma_0=\gamma_1=\Gamma/8$ and $\gamma_2=3\Gamma/4$, where $\Gamma=1/\tau$ is the total relaxation rate of Rydberg state with $\tau$ being the Rydberg lifetime. In the system of laser-driven natural atoms, the dephasing rate is in general much less than the relaxation rate~\cite{Levine2018,Omran2019,Levine2019}, and then here we set $\gamma_{\phi j}/2\pi=1$~kHz.

Here we take the two- and three-qubit holonomic gates as examples, and show the average fidelities to verify the arbitrariness of an initial state. The average fidelity is defined based on a trace-preserving operator, as~\cite{Nielsen2002}
\begin{eqnarray}
	\bar{F}(\varepsilon, \hat{U})=\frac{\sum_{v=1}^{4^{N+1}} \operatorname{tr}\left[\hat{U} \hat{u}_{v}^{\dagger} \hat{U}^{\dagger} \varepsilon\left(\hat{u}_{v}\right)\right]+l^{2}}{l^{2}(l+1)},
\end{eqnarray}
where $\hat{u}_v=\bigotimes_k^{N+1}{\hat\sigma}_k$ is a tensor product of Pauli matrices ${\hat\sigma}_k\in\{\hat I, \hat\sigma^x, \hat\sigma^y, \hat\sigma^z\}$ on computational states $\{|0\rangle,~|1\rangle\}$, and $l=2^{N+1}$ for an $(N+1)$-qubit gate. $\varepsilon(\hat{u}_{v})$ is a trace-preserving quantum operation obtained through solving the master equation.
We show average fidelities of various two- and three-qubit controlled-$\hat{U}_{t}(\theta,\phi,\gamma)$ gates in Figs.~\ref{f11}{(a)} and {(b)}, respectively. \ The evolutionary trends of the average fidelity for different gates are dependent of the value of $\gamma$, and the two- and three-qubit gates can be achieved with average fidelities over $0.995$.

\end{document}